%% file: main.tex
\documentclass[journal]{IEEEtran}
\usepackage{amsmath,amsthm,amssymb,amsfonts}
\usepackage{graphicx}
\usepackage{color}
\usepackage{subfigure}
\usepackage{balance}
\usepackage{hyperref}
\usepackage{textcomp}
\usepackage{xcolor}
\usepackage{cite}
\usepackage{multicol,lipsum}
\usepackage[utf8x]{inputenc}
\usepackage{epsfig}
\usepackage{algorithm}
\usepackage{algorithmic}
\usepackage{multirow}
\usepackage{hyperref}
\usepackage{mathtools}

\usepackage{bm}
\usepackage{listings}
\newtheorem{remark}{Remark}
\newtheorem{definition}{Definition}
\newtheorem{theorem}{Theorem}
\newtheorem{proposition}{Proposition}

\newtheorem{example}{Example}

\newtheorem{assumption}{Assumption}
\newtheorem{problem}{Problem}
\newcommand{\oomit}[1]{}

\begin{document}
\title{Reach-avoid Controllers Synthesis for Safety Critical Systems}
\author{Bai Xue\\
State Key Lab. of Computer Science, Institute of Software, CAS, Beijing, China\\
\{xuebai@ios.ac.cn\} 
}

\maketitle

\begin{abstract}
In this paper we propose sufficient conditions to synthesizing reach-avoid controllers for deterministic systems modelled by ordinary differential equations and stochastic systems modeled by stochastic differential equations  based on the notion of control guidance-barrier functions. We begin with considering deterministic systems. Given an open safe set, a target set and a nominal controller, we attempt to synthesize a reach-avoid controller, which modifies the nominal controller in a minimal way enforcing the reach-avoid objective, i.e.,  the system will enter the target set  eventually while staying inside the safe set before the first target hitting time. Three control guidance-barrier functions are developed and three corresponding conditions for synthesizing reach-avoid controllers are constructed, which get progressively weaker and thus facilitate the optimal controller design. The first and second ones are termed exponential and asymptotic control guidance-barrier functions, which guarantee that every trajectory starting from the safe set will enter the target set respectively at an exponential rate and in an asymptotic way. However, it is observed that these two control guidance-barrier functions have one condition in common of enforcing invariance of its every positive sublevel set until the system enters the target set, which is strict and thus limits the space of admissible controllers. Consequently, a lax control guidance-barrier function is further developed such that only the safe set set is an invariance before the system enters the target set, expanding the space of admissible control inputs. Then, we extend the methodologies for deterministic systems to stochastic systems, which synthesize reach-avoid controllers  in the probabilistic setting. Finally, several numerical examples demonstrate the performance of proposed methods.
\end{abstract}

\input{introduction}

\input{preliminaries}

\input{controllers}

\input{stochastic}

\input{example}
\input{conclusion}
\bibliographystyle{abbrv}
\bibliography{reference}
\input{appendix}

\end{document}

%% file: introduction.tex
\section{Introduction}
\label{intro}
Cyber-physical systems are becoming ubiquitous due to rapid advances in computation, communication, and memory \cite{rajkumar2010cyber}, and can be found in diverse application domains including automobiles, aviation, advanced manufacturing, and integrated medical devices \cite{baheti2011cyber}. Many of these systems  are safety-critical and thus must satisfy safety requirements, which are typically formulated as forward invariance of a given safe set \cite{ames2019control}, on their dynamic trajectories in order to  prevent economic harm and loss of life. The need for safety has motivated extensive research into verifying and synthesizing controllers to satisfy safety requirements.

 Common techniques include Lyapunov and barrier methods \cite{prajna2004safety}, discrete approximations, and direct computations of reachable sets \cite{kurzhanski2000ellipsoidal,althoff2021set}, among others, have been developed for safety verification of safety-critical systems with given controllers. However, as the systems being verified get more complicated, the traditionally verified and tuning/redesigning (``design V") iteration procedure \cite{forsberg1991relationship} becomes harder and more time-consuming. Consequently, there has been increasing interest in synthesizing controllers with provable safety guarantees in the design phase, and thus many methods emerge such as control barrier functions methods \cite{ames2019control}, Hamilton-Jacobi methods \cite{tomlin1998synthesizing}, moment based optimization methods \cite{majumdar2014convex}, and controlled invariant sets methods \cite{blanchini1999set,li2017invariance}. Such methods are instrumental in removing  bad controllers at an early state and avoiding the time-consuming ``design V" iteration, consequently reducing software development expenses. Besides safety objectives, cyber-physical systems such as autonomous vehicles, industrial robots, and chemical reactors often require simultaneous satisfaction of performance specifications. For instance, \cite{ames2016control} considers the design of controllers enforcing objective of joint safety and stability. Among the many provably correct controllers synthesis problems, there is a special one of synthesizing reach-avoid controllers. Reach-avoid controllers enforce both safety via avoiding a set of unsafe states and reachability via reaching a desired target set. Consequently, they can address many important engineering problems such as collision avoidance  \cite{margellos2010hamilton} and target surveillance, and thus have turned out to be of fundamental importance in engineering.   

In this paper we investigate the reach-avoid controllers synthesis problem for continuous-time systems modelled by ordinary differential equations and stochastic differential equations. A reach-avoid controller is a controller such that the system starting from a specified safe set will eventually enter a desired target set while staying inside the safe set before the first target hitting time, in a deterministic sense for deterministic systems or in a probabilistic sense for stochastic systems. For deterministic systems, three (exponential/asymptotic/lax) control guidance-barrier functions are developed for synthesizing reach-avoid control inputs. These three functions generate three increasingly-weak sufficient conditions, which expand the space of admissible control inputs gradually. The first and second ones are termed exponential and asymptotic control guidance-barrier functions, which guarantee that every trajectory starting from the safe set will respectively enter the target set at an exponential rate and in an asymptotic way. However, it is observed that these two control guidance-barrier functions have one condition in common of enforcing invariance of its every positive sublevel set until the system enters the target set, which is rather restrict and thus limits the space of admissible controllers. Consequently, a lax control guidance-barrier function is developed and a weaker condition is obtained which does not enforce invariance of its every positive sublevel set, thus expanding the space of admissible controllers. The first two control guidance-barrier functions are then extended to stochastic systems, which is to synthesize a reach-avoid controller in a minimal fashion such that the system enters a target set eventually while staying inside a safe set before the first target hitting time in probability. Finally, several numerical examples demonstrate the theoretical developments of  proposed methods. The contributions of this work are summarized below. 
\begin{enumerate}
    \item A unified framework of synthesizing reach-avoid controllers for both deterministic systems modelled by ordinary differential equations and stochastic systems modelled by stochastic differential equations in an optimal sense is developed. 
   \item Within this framework, several control guidance-barrier functions are respectively proposed for constructing sufficient conditions to synthesize reach-avoid controllers, for both deterministic systems and stochastic systems. These three conditions are becoming less restrictive gradually. Consequently, an increasingly large set of control inputs compatible with reach-avoid specifications will be obtained, which either provides more chances for synthesizing a reach-avoid controller successfully or improves the already obtained controller.
\end{enumerate}

\subsection*{Related Work}
Multiple solution techniques have been proposed for safety-critical control synthesis for cyber-physical systems, including Hamilton-Jacobi-Bellman-Isaacs(HJI) equation \cite{tomlin1998conflict}, mixed-integer program \cite{mellinger2012mixed}, viability theory \cite{aubin2011viability},  control barrier function,  and control Lyapunov function -based methodologies \cite{ames2019control}. We do not intend to give a comprehensive review of related literature herein, but instead introduce the closest related works, which mainly center around control barrier function based methodologies.  Control barrier functions were inspired by barrier certificates, which were originally proposed for safety verification of continuous-time deterministic and stochastic systems \cite{prajna2004safety,prajna2007framework}. The control barrier functions framework has several advantages in addition to provable safety guarantees. For affine control systems, control barrier functions lead to linear constraints on the control, which can be used to design computationally tractable optimization-based controllers. They can also be easily composed with control Lyapunov functions to provide joint guarantees on stability and safety.

For deterministic systems, the notion of a barrier certificate was first extended to a control version in \cite{wieland2007constructive}. This can be thought of as being analogous to how \cite{artstein1983stabilization,sontag1989universal} extended Lyapunov functions to control Lyapunov functions in order to enable controller synthesis for stabilization tasks. Later on, various control barrier certificates such as reciprocal control barrier functions \cite{ames2014control}, zeroing control barrier functions \cite{ames2016control}, and minimal control barrier functions \cite{konda2020characterizing} have been developed to design controllers for enforcing safety. These control barrier certificates mainly differ in their expressiveness.  To address safety constraints with higher relative-degree, the method of control barrier functions was extended to position-based constraints with relative degree 2 in \cite{wu2015safety,nguyen2015safety}, and then  \cite{ames2014control} extended it to arbitrary high relative-degree systems using a backstepping based method. Recently,  exponential control barrier functions were proposed in \cite{nguyen2016exponential} to enforce high relative-degree safety constraints.

In addition to safety,  the control design of safety-critical dynamical systems in practice  often requires simultaneous satisfaction of performance specifications and safety constraints. The problem of safe stabilization, i.e., to stabilize the system while maintaining the system to stay within a given safe set, attracts wide attentions recently. In order to simultaneously achieve safety and stabilization of dynamical systems, a number of control design methods have been proposed in the literature to compose control Lyapunov functions with control barrier functions. The unification of control Lyapunov functions and control barrier functions appeared in \cite{romdlony2014uniting,romdlony2016stabilization,ames2014control,ames2016control}. The objective of the works \cite{romdlony2014uniting,romdlony2016stabilization} was to incorporate into a single feedback law the conditions required to simultaneously achieve asymptotic stability of an equilibrium point, while avoiding an unsafe set. In contrast, the approach of the works \cite{ames2014control,ames2016control} was to pose a feedback design problem that mediates safety and stabilization requirements, in the sense that safety is always guaranteed, and progress toward the stabilization objective is assured when the two requirements are not in conflict. Meanwhile, based on the combination of exponential control barrier functions and control Lyapunov functions, \cite{nguyen2016exponential} studied the safety via enforcing high relative-degree safety-critical constraints and stability.  \cite{jankovic2018robust} further considered the relaxation on the lie derivative of control Lyapunov functions such that local asymptotic stability can be guaranteed. The present work also investigates the controllers synthesis problem for enforcing simultaneous satisfaction of joint safety and performance  objectives. However, it is different from the above mentioned works. The differences are twofold: the present work studies the problem of synthesizing reach-avoid controllers, which enforce safety via staying inside a safe set and performance via reaching a desired target set rather than approaching an equilibrium asymptotically. When the target set is a region of attraction, the synthesized reach-avoid controller is also able to enforce safety and stability; the sufficient conditions developed in this work rely on the notion of control guidance-barrier functions, which are a control version of guidance-barrier functions proposed in \cite{xue2022_reach_verification},  and these functions are not  a combination of control barrier functions and control Lyapunov functions.  In \cite{xue2022_reach_verification} exponential and asymptotic guidance-barrier functions were proposed for computing inner-approximations of reach-avoid sets for deterministic systems. Although exponential and asymptotic control guidance-barrier functions are a direct extension of the guidance-barrier functions in \cite{xue2022_reach_verification}, the present work further proposes lax control guidance-barrier functions, which are weaker than the former two control  functions.

Compared to deterministic systems, literature on control barrier functions for systems with stochastic disturbances is very scarce. When considering stochastic systems, i.e., dynamical systems involving stochastic processes, solving the controllers synthesis problem qualitatively in a non-stochastic manner usually gives pessimistic answers, since generally resultant bounds on the values of stochastic inputs will be overly conservative. It is  natural to formulate and solve probabilistic variants of controllers synthesis problems. For instance, it requires the synthesis of $p$-controllers to guarantee with probability being larger than $p$ that the dynamical system satisfies the safety requirement. \cite{tamba2021notion} examined the safety verification of stochastic systems using the notion of stochastic zeroing barrier functions. The first extension of control barrier functions for deterministic systems to stochastic systems appeared in  \cite{clark2019control}, which proposed stochastic reciprocal control barrier functions and derived sufficient conditions for the stochastic system to satisfy safety constraints with probability one. Motivated by the need to reduce potentially severe control constraints generated by stochastic reciprocal control barrier functions in the neighborhood of the safety boundary, \cite{wang2021safety} proposed a notion of stochastic control barrier functions for safety-critical control by providing the worst-case probability estimation. Recently, \cite{clark2021control} extended the results in \cite{clark2019control} by introducing stochastic zeroing control barrier functions constructions that guarantee safety with probability one. It also proposed control policies that ensure safety and stability by solving quadratic programs containing control barrier functions and stochastic control Lyapunov functions. In this work we extend the aforementioned exponential and asymptotic control guidance-barrier functions for deterministic systems to stochastic systems for synthesizing reach-avoid controllers such that the system enters a target set eventually while staying inside a specified safe set before the first target hitting time in probability.  Among these two extensions, the asymptotic control guidance-barrier function  is constructed based on guidance-barrier functions in \cite{xue2022_Stochastic}, which was developed to inner-approximating reach-avoid sets for stochastic systems.  

The structure of this paper is organized as follows. Three control guidance-barrier functions  are proposed in  Section \ref{OTRACS}, which also elucidates three sufficient conditions for synthesizing reach-avoid controllers of deterministic systems modelled by ordinary differential equations. Section \ref{RCSSS} extends some control guidance-barrier functions in Section \ref{OTRACS} to stochastic systems modelled by stochastic differential equations and proposes two sufficient conditions for synthesizing reach-avoid controllers in the probabilistic setting. 

The following  notations will be used throughout the rest of this paper: $\mathbb{R}^n$ denotes
the set of n-dimensional real vectors; $\mathbb{R}_{\geq 0}$ denotes non-negative real numbers; the closure of a set $\mathcal{X}$ is denoted by $\overline{\mathcal{X}}$, the complement by $\mathcal{X}^c$ and the boundary by $\partial \mathcal{X}$; $\|\bm{x}\|$ denotes the 2-norm, i.e., $\|\bm{x}\|:=\sqrt{\sum_{i=1}^n x_i^2}$, where $\bm{x}=(x_1,\ldots,x_n)^{\top}$; vectors are denoted by boldface letters.  For a vector function $\bm{g}(\bm{x}): \mathcal{C}\rightarrow \mathbb{R}^m$, $\|\bm{g}(\bm{x})\|=\sqrt{\int_{\bm{x}\in \mathcal{C}}\bm{g}^{\top}(\bm{x})\bm{g}(\bm{x})d\bm{x}}$.

%% file: preliminaries.tex
\section{Reach-avoid Controllers Synthesis for Deterministic Systems}
\label{RACS}
In this section we focus our attention on solving the reach-avoid controllers synthesis problem for deterministic systems modelled by ordinary differential equations.

\subsection{Preliminaries}
\label{CBF}
In this section we give an introduction on the system and reach-avoid controllers synthesis problem of interest. 

Consider an affine control system,
\begin{equation}
    \label{control_system}
    \dot{\bm{x}}=\bm{f}(\bm{x})+\bm{g}(\bm{x})\bm{u},
\end{equation}
with $\bm{f}$ and $\bm{g}$ locally Lipschitz, $\bm{x}\in \mathbb{R}^n$ and $\bm{u}\in \mathcal{U}\subseteq \mathbb{R}^m$ is the set of admissible control inputs.

Let $\bm{u}(\bm{x}): \mathbb{R}^n\rightarrow \mathcal{U}$ be a feedback controller such that the resulting system \eqref{control_system} is locally Lipschitz. Then for any initial condition $\bm{x}_0:=\bm{x}(0)\in \mathbb{R}^n$, there exists a maximum time interval $I(\bm{x}_0,\bm{u})=[0,\tau_{\max})$ such that $\bm{\phi}_{\bm{x}_0}(\cdot): I(\bm{x}_0,\bm{u})\rightarrow \mathbb{R}^n$ is the unique solution to system \eqref{control_system}, where $\bm{\phi}_{\bm{x}_0}(0)=\bm{x}_0$ and $\tau_{\max}$ is the explosion time with \[\lim_{t\rightarrow  \tau_{\max}}\bm{\phi}_{\bm{x}_0}(t)=\infty. \] 

Given a bounded open safe set $\mathcal{C}$ and a target set $\mathcal{X}_r$, the reach-avoid property with respect to them and associated reach-avoid controllers are formulated in Definition \ref{RAC}. 

\begin{definition}[Reach-avoid Controllers]
\label{RAC}
Given a feedback controller $\bm{u}(\cdot):\mathbb{R}^n\rightarrow \mathcal{U}$, the reach-avoid property with respect to the safe set $\mathcal{C}$ and target set $\mathcal{X}_r$ is satisfied if, starting from any initial state in $\mathcal{C}$, system \eqref{control_system} with this controller will enter the target set $\mathcal{X}_r$ eventually while staying inside $\mathcal{C}$ before the first target hitting time.  If $\bm{u}(\cdot):\mathbb{R}^n \rightarrow \mathcal{U}$ is locally Lipschitz continuous, it is a reach-avoid controller with respect to the safe set $\mathcal{C}$ and target set $\mathcal{X}_r$.
 \end{definition}
 
 Then, we formulate our  reach-avoid controllers synthesis problem.
 \begin{problem}[Reach-avoid Controllers Synthesis]
 \label{problem1}
Given a feedback control $\bm{u}=\bm{k}(\bm{x})$ for the control system \eqref{control_system}, which may not be a reach-avoid controller, we wish to synthesize a reach-avoid controller in a minimally invasive fashion, i.e., modify the existing controller $\bm{k}(\bm{x})$ in a minimal way so as to guarantee satisfaction of the reach-avoid specification via solving the following optimization:
\begin{equation}
    \label{q_P0}
    \begin{split}
        &\bm{u}(\bm{x})=\arg\min \|\bm{u}(\bm{x})-\bm{k}(\bm{x})\| \\
        &s.t.~\bm{u}(\cdot): \mathcal{C}\rightarrow \mathcal{U} \text{~is a reach-avoid controller.}  
     \end{split}
\end{equation}
\end{problem}

Besides, we impose the following assumptions on the safe set $\mathcal{C}$ and target set $\mathcal{X}_r$.

\begin{assumption}
\label{assump}
The safe set $\mathcal{C}$ and target set $\mathcal{X}_r$ satisfy the following conditions:
\begin{enumerate}
    \item $\mathcal{C}=\{\bm{x}\in \mathbb{R}^n\mid h(\bm{x})>0\}$,
    \item $\overline{\mathcal{C}}=\{\bm{x}\in \mathbb{R}^n \mid  h(\bm{x})\geq 0\}$,
    \item $\partial \mathcal{C}=\{\bm{x}\in \mathbb{R}^n\mid h(\bm{x})=0\}$,
\end{enumerate}
where $h(\cdot):\mathbb{R}^n \rightarrow \mathbb{R}$ is a continuously differentiable function, and 
\begin{enumerate}
    \item $\mathcal{C}$ has no isolated point, 
\item $\mathcal{C}\cap \mathcal{X}_r$ is not empty and has no isolated point.
\end{enumerate}
\end{assumption}

Although this work studies the reach-avoid controllers synthesis problem, the proposed methods can also be directly extended to the synthesis of controllers enforcing safety and stability as in \cite{ames2016control}, with the assumption that the target set $\mathcal{X}_r$ is a region of attraction. 

\begin{remark}
In this paper we focus our attention on the assumption that the safe set $\mathcal{C}$ is open. Some of our methods proposed in this paper also work with a compact set $\mathcal{C}$. We will remark this extension in an appropriate place in the sequel. 
\end{remark}

%% file: controllers.tex
\subsection{Reach-avoid Controllers Synthesis}
\label{OTRACS}
In this section we elucidate our methods on synthesizing reach-avoid controllers by solving optimization \eqref{q_P0}. The methods are built upon three control (i.e., exponential, asymptotic and lax) guidance-barrier functions. 

\subsubsection{Exponential Control Guidance-barrier Functions}
\label{egbf}
In this subsection we introduce exponential control guidance-barrier functions for the synthesis of reach-avoid controllers. This function is derived from the exponential guidance-barrier function in \cite{xue2022_reach_verification}.

\begin{definition}
Given the safe set $\mathcal{C}$ and tareget set $\mathcal{X}_r$ satisfying Assumption \ref{assump}, $h(\cdot):\mathbb{R}^n \rightarrow \mathbb{R}$ is an exponential control guidance-barrier function if there exists $\lambda>0$ such that
\begin{equation*}
\sup_{\bm{u}(\bm{x})\in \mathcal{U}}\mathcal{L}_{h,\bm{u}}(\bm{x})\geq \lambda h(\bm{x}), \forall \bm{x}\in \overline{\mathcal{C}\setminus \mathcal{X}_r},
\end{equation*}
where $\mathcal{L}_{h,\bm{u}}(\bm{x})=\bigtriangledown_{\bm{x}}h(\bm{x})\cdot \bm{f}(\bm{x})+\bigtriangledown_{\bm{x}} h(\bm{x})\cdot \bm{g}(\bm{x})\bm{u}(\bm{x})$. $\hfill\blacksquare$ 
\end{definition}

The exponential control guidance-barrier function will facilitate the construction of constraints for synthesizing reach-avoid controllers, which are formulated in Theorem \ref{exponential_control_pro}.

\begin{theorem}
\label{exponential_control_pro}
Given the safe set $\mathcal{C}$ and target set $\mathcal{X}_r$  satisfying Assumption \ref{assump}, if the function $h(\cdot): \mathbb{R}^n \rightarrow \mathbb{R}$ is an exponential control guidance-barrier function, then any Lipschitz continuous controller $\bm{u}(\cdot):\mathcal{C}\rightarrow \mathcal{U}$ such that $\bm{u}(\bm{x})\in \mathcal{K}_e(\bm{x})$  is a reach-avoid controller with respect to the safe set $\mathcal{C}$ and target set $\mathcal{X}_r$, where 
\[
\mathcal{K}_e(\bm{x})=\{\bm{u}(\bm{x})\in \mathcal{U}\mid \text{constraint}~\eqref{exponential_control}~\text{holds}\}
\]
with 
\begin{equation}
\label{exponential_control}
\begin{cases}
    \mathcal{L}_{h,\bm{u}}(\bm{x})-\lambda h(\bm{x})\geq 0, \forall \bm{x}\in \overline{\mathcal{C}\setminus \mathcal{X}_r}, \\
    \lambda>0.
    \end{cases}
\end{equation}
\end{theorem}

\begin{remark}
\label{remark1}
When the safe set $\mathcal{C}$ is compact, i.e., $\mathcal{C}=\{\bm{x}\in \mathbb{R}^n \mid h(\bm{x})\geq 0\}$, a slight modification to constraint \eqref{exponential_control} is to replace  the `greater than or equal to sign' with the `greater sign' in the first condition $ \mathcal{L}_{h,\bm{u}}(\bm{x})-\lambda h(\bm{x})\geq 0, \forall \bm{x}\in \overline{\mathcal{C}\setminus \mathcal{X}_r}$, i.e., 
\begin{equation*}
  \mathcal{L}_{h,\bm{u}}(\bm{x})-\lambda h(\bm{x})>0, \forall \bm{x}\in \overline{\mathcal{C}\setminus \mathcal{X}_r},
\end{equation*}
where $\lambda>0$. This modification is to enforce the reachability objective for states on the boundary $\partial \mathcal{C}$.   $\hfill\blacksquare$
\end{remark}

\begin{remark}
\label{smaller}
    It is observed that the smaller $\lambda$ is, the weaker constraint \eqref{exponential_control} is. However, $\lambda$ cannot be zero. 
It is also observed that an exponential control guidance-barrier function complements a zeroing control barrier function satisfying \[
\begin{cases}
&\sup_{\bm{u}(\bm{x})\in \mathcal{U}}\mathcal{L}_{h,\bm{u}}(\bm{x})\geq \lambda h(\bm{x}), \forall \bm{x}\in \overline{\mathcal{C}},\\
&\lambda<0.
\end{cases}\]
Such a zeroing control barrier function facilitates the design of controllers only for enforcing safety.
\end{remark}

 The set $\mathcal{K}_e$ is convex, thus a direct computation of a locally Lipschitz continuous feedback controller $\bm{u}(\bm{x})$ satisfying constraint \eqref{exponential_control} using convex optimization is possible. Based on the exponential guidance-barrier function $h(\bm{x})$ and Theorem \ref{exponential_control_pro},  Problem \ref{problem1} could be addressed via solving the following  program \eqref{qu}.
\begin{equation} 
\label{qu}
\begin{split}
        &\min_{\bm{u}(\bm{x})\in \mathcal{U},\lambda,\delta} \|\bm{u}(\bm{x})-\bm{k}(\bm{x})\|+c\delta\\
s.t.~&\mathcal{L}_{h,\bm{u}}(\bm{x})-\lambda h(\bm{x})\geq -\delta, \forall \bm{x}\in \overline{\mathcal{C}\setminus \mathcal{X}_r};\\
&\delta\geq 0;\\
&\lambda\geq \xi_0,
        \end{split}
\end{equation}
where $\bm{u}(\bm{x}): \overline{\mathcal{C}}\rightarrow \mathcal{U}$ is a locally Lipschitz parameterized function with some unknown parameters, $\xi_0$ is a user-defined positive value which is to enforce the strict positivity of $\lambda$, and $c$ is a large constant that penalizes safety/reachability violations.  As commented in Remark \ref{smaller}, the smaller $\xi_0$ is, the weaker the constraint in \eqref{qu} is.  However, $\xi_0$ cannot be zero. Thus, in numerical computations, we do not recommend the use of too small $\xi_0$ due to numerical errors. The optimization is not sensitive to the $c$ value as long as it is very large (e.g., $10^{12}$), such that the constraint
violations are heavily penalized. When $\delta>0$, it provides graceful degradation  when the reach-avoid objective cannot be enforced.  Via solving optimization \eqref{qu}, we have the following conclusion.
\begin{proposition}
\label{deltadayu0_ex}
        Suppose $(\bm{u}_0(\bm{x}),\lambda,\delta)$ is obtained by solving optimization \eqref{qu}, and $\bm{u}_0(\bm{x})$ is locally Lipschitz continuous, 
        \begin{enumerate}
            \item If $\delta=0$, $\bm{u}_0(\bm{x})$ is a reach-avoid controller with respect to the safe set $\mathcal{C}$ and target set $\mathcal{X}_r$.
            \item If $\delta>0$ and $\mathcal{D}\cap \mathcal{X}_r\neq \emptyset$, where $\mathcal{D}=\{\bm{x}\mid h(\bm{x})>\frac{\delta}{\lambda}\}$,  $\bm{u}_0(\bm{x})$ is the reach-avoid controller with respect to the set $\mathcal{D}$ and target set $\mathcal{X}_r$.
        \end{enumerate}
\end{proposition}

If $\bm{u}(\bm{x}) \in \mathcal{K}_e(\bm{x})$, it will drive system \eqref{control_system} to enter the target set $\mathcal{X}_r$ at an exponential rate of $\lambda$. Besides, it requires $h(\bm{\phi}_{\bm{x}_0}(t))$ to be strictly monotonically increasing with respect to $t$ before the trajectory $\bm{\phi}_{\bm{x}_0}(t)$ enters the target set $\mathcal{X}_r$. Thus, if there exists $\bm{y}\in \mathcal{C}\setminus \mathcal{X}_r$ such that $h(\bm{y})\geq \sup_{\bm{x}\in \mathcal{X}_r}h(\bm{x})$, we cannot obtain a reach-avoid controller with respect to the safe set $\mathcal{C}$ and the target set $\mathcal{X}_r$ via solving \eqref{qu}.
Therefore,  solving \eqref{qu} may either obtain a pessimistic reach-avoid controller or lead to $\delta>0$. In the following we will  obtain a new optimization with a set of weaker constraints for synthesizing reach-avoid controllers.

\subsubsection{Asymptotic Control Guidance-barrier Functions}
\label{ACGBF}
In this subsection we introduce asymptotic control guidance-barrier functions for the synthesis of reach-avoid controllers. The asymptotic control guidance-barrier function is derived from the asymptotic guidance-barrier function in \cite{xue2022_reach_verification}, which corresponds to the case that $\lambda=0$ in constraint \eqref{exponential_control}.
 
\begin{definition}
Given the safe set $\mathcal{C}$ and target set $\mathcal{X}_r$ satisfying Assumption \ref{assump}, then $h(\cdot):\mathbb{R}^n \rightarrow \mathbb{R}$ is called an asymptotic control guidance-barrier function, if there exists a continuously differentiable function $w(\cdot): \mathbb{R}^n\rightarrow \mathbb{R}$ such that 
\begin{equation*}
    \begin{split}
\sup_{\bm{u}(\bm{x})\in \mathcal{U}} \min\big\{ \mathcal{L}_{h,\bm{u}}(\bm{x}), \mathcal{L}_{w,\bm{u}}(\bm{x})-h(\bm{x})\big\}\geq 0, \forall \bm{x}\in \overline{\mathcal{C}\setminus \mathcal{X}_r}.  \hfill\blacksquare
    \end{split}
\end{equation*}
\end{definition}

The asymptotic control guidance-barrier function also facilitates the construction of constraints for synthesizing reach-avoid controllers, which are formulated in Theorem \ref{asymp_control}.
\begin{theorem}
\label{asymp_control}
Given the safe set $\mathcal{C}$ and target set $\mathcal{X}_r$ satisfying Assumption \ref{assump}, if the function $h(\cdot): \mathbb{R}^n \rightarrow \mathbb{R}$ is an asymptotic control guidance-barrier function, then any Lipschitz continuous controller $\bm{u}(\cdot):\mathcal{C}\rightarrow \mathcal{U}$ such that $\bm{u}(\bm{x})\in \mathcal{K}_a(\bm{x})$ is a reach-avoid controller with respect to the safe set $\mathcal{C}$ and target set $\mathcal{X}_r$, where 
\begin{equation*}
\mathcal{K}_a(\bm{x})=\left\{\bm{u}(\bm{x})\in \mathcal{U} \middle|\; \text{constraint}~\eqref{control_g_b_f}~\text{holds}
\right
\}
\end{equation*}
with  
\begin{equation}
\label{control_g_b_f}
\begin{cases}
&\mathcal{L}_{h,\bm{u}}(\bm{x})\geq 0,  \forall \bm{x}\in \overline{\mathcal{C}\setminus \mathcal{X}_r}, \\
&\mathcal{L}_{w,\bm{u}}(\bm{x})-h(\bm{x})\geq 0, \forall \bm{x}\in \overline{\mathcal{C}\setminus \mathcal{X}_r}.
\end{cases}
\end{equation}
\end{theorem}

According to Theorem \ref{asymp_control}, Problem \ref{problem1} could be addressed via solving the following optimization \eqref{qua_asym0}:
\begin{equation} 
\label{qua_asym0}
\begin{split}
        &\min_{\bm{u}(\bm{x})\in \mathcal{U},w(\bm{x}),\delta} \|\bm{u}(\bm{x})-\bm{k}(\bm{x})\|+c\delta\\
s.t.~&\mathcal{L}_{h,\bm{u}}(\bm{x})\geq 0,   \forall \bm{x}\in \overline{\mathcal{C}\setminus \mathcal{X}_r},\\
&\mathcal{L}_{w,\bm{u}}(\bm{x})-h(\bm{x})\geq -\delta, \forall \bm{x}\in \overline{\mathcal{C}\setminus \mathcal{X}_r},\\
&\delta\geq 0,
        \end{split}
\end{equation}
where $\bm{u}(\bm{x}): \overline{\mathcal{C}}\rightarrow \mathcal{U}$ is a locally Lipschitz parameterized function with some unknown paramteres, and $c$ is a large constant that penalizes reachability violations.

Via comparing constraints \eqref{control_g_b_f} and \eqref{exponential_control}, it is easy to observe that constraint \eqref{control_g_b_f} is more expressive and thus admits more feasible feedback controllers, since constraint \eqref{control_g_b_f} will be reduced to \eqref{exponential_control} when $w(\bm{x})=\frac{h(\bm{x})}{\lambda}$. Consequently, it is more likely to either obtain a reach-avoid controller with respect to the safe set $\mathcal{C}$ and target set $\mathcal{X}_r$, or improve the obtained  reach-avoid controller when using constraint \eqref{control_g_b_f} for computations and thus lead to smaller $\|\bm{u}(\bm{x})-\bm{k}(\bm{x})\|$. Besides, comapred to exponential control guidance-barrier functions, a reach-avoid controller generated by  an asymptotic control guidance-barrier function requires $h(\bm{\phi}_{\bm{x}_0}(t))$ to be monotonically non-decreasing with respect to $t$ rather than strictly monotonically increasing before the trajectory $\bm{\phi}_{\bm{x}_0}(t)$ enters the target set $\mathcal{X}_r$, thereby diversifying the application scenarios. However, the synthesis of a reach-avoid controller $\bm{u}(\bm{x})$ via solving constraint \eqref{control_g_b_f} is a nonlinear problem due to the existence of the term $\bigtriangledown_{\bm{x}}w(\bm{x})\cdot g(\bm{x})\bm{u}$. We can adopt an iterative algorithm to solve optimization \eqref{qua_asym0} by decomposing it into two convex sub-problems based on two cases, which correspond to whether a reach-avoid controller with respect to the safe set $\mathcal{C}$ and target set $\mathcal{X}_r$ is obtained via solving optimization \eqref{qua_asym0} or not.

The iterative algorithm for solving optimization \eqref{qua_asym0} is elucidated in the following.

\textbf{Case 1: a reach-avoid controller is returned via solving optimization \eqref{qu}, that is, $\delta=0$ in \eqref{qu}.}
According to the comparison between constraints \eqref{control_g_b_f} and \eqref{exponential_control} above, a reach-avoid controller $\bm{u}_0(\bm{x})$ from optimization \eqref{qu} is also a solution to optimization \eqref{qua_asym0}. Consequently, we can adopt the iterative algorithm in Alg. \ref{alg_0} to solve optimization \eqref{qua_asym0} for improving the controller $\bm{u}_0(\bm{x})$.

\begin{algorithm}
\caption{An iterative procedure for solving optimization \eqref{qua_asym0} when a reach-avoid controller is returned via solving optimization \eqref{qu}.}
 Let $\bm{u}_0(\cdot): \mathbb{R}^n \rightarrow \mathcal{U}$ be computed via solving optimization \eqref{qu}, and
$\epsilon'>0$ be a specified  threshold.
\begin{algorithmic}
\STATE $\xi_0:=\|\bm{u}_0(\bm{x})-\bm{k}(\bm{x})\| $;
\STATE $i:=0$;
\WHILE{TRUE}
    \STATE compute $w_i^*(\bm{x})$ such that 
    \[ \mathcal{L}_{w_i,\bm{u}_i}(\bm{x})-h(\bm{x})\geq 0, \forall \bm{x}\in \overline{\mathcal{C}\setminus \mathcal{X}_r};\]
    \STATE solve the following program to obtain $\bm{u}_{i+1}(\bm{x})$:
    \begin{equation*} 
\begin{split}
        &\min_{\bm{u}(\bm{x})\in \mathcal{U}} \|\bm{u}(\bm{x})-\bm{k}(\bm{x})\|\\
s.t.~&\mathcal{L}_{h,\bm{u}}(\bm{x})\geq 0, \forall \bm{x}\in \overline{\mathcal{C}\setminus \mathcal{X}_r};\\
&\mathcal{L}_{w_i^*,\bm{u}}(\bm{x})-h(\bm{x})\geq 0, \forall \bm{x}\in \overline{\mathcal{C}\setminus \mathcal{X}_r};
        \end{split}
\end{equation*}
    \IF{$\xi_{i+1}-\xi_i\leq -\epsilon'$, where $\xi_{i+1}=\|\bm{u}_{i+1}(\bm{x})-\bm{k}(\bm{x})\|$}
    \STATE  $i:=i+1$;
    \ELSE
    \STATE return $\bm{u}_{i+1}(\bm{x})$ and terminate;
    \ENDIF
    \ENDWHILE
\end{algorithmic}
\label{alg_0}
\end{algorithm}

\textbf{Case 2: a reach-avoid controller is not returned via solving optimization \eqref{qu}, that is, $\delta>0$ in \eqref{qu}.}
When a reach-avoid controller $\bm{u}(\bm{x})$ is not computed via solving optimization \eqref{qu}, one reason leading to this failure is due to the strictness of constraint \eqref{exponential_control} and the other one is the potential conflict between reachability and safety. At all events,  in safety-critical systems the safety has the highest priority. 

One of advantages of constraint \eqref{control_g_b_f} over \eqref{exponential_control} lies in its ability of characterizing safety and reachability separately. The condition 
\begin{equation}
\label{asymp_safety}
\mathcal{L}_{h,\bm{u}}(\bm{x})\geq 0, \forall \bm{x}\in \overline{\mathcal{C}\setminus \mathcal{X}_r}
\end{equation}
ensures safety, i.e., trajectories starting from $\mathcal{C}\setminus \mathcal{X}_r$ will either stay inside the set $\mathcal{C}\setminus \mathcal{X}_r$ for all the time or enter the target set $\mathcal{X}_r$ eventually while staying inside the safe set $\mathcal{C}$ before the first target hitting time.  We term the Lipschitz controller $\bm{u}(\bm{x})$ satisfying constraint \eqref{asymp_safety} a safe controller.
On the other hand, the condition 
\begin{equation}
\label{asymp_performance}
    \mathcal{L}_{w,\bm{u}}(\bm{x})-h(\bm{x})\geq 0,\forall \bm{x}\in \overline{ \mathcal{C}\setminus \mathcal{X}_r}
\end{equation}
ensures reachability, i.e., trajectories starting from $\mathcal{C}\setminus \mathcal{X}_r$ will eventually enter the target set $\mathcal{X}_r$.  It is worth remarking that a function $w(\bm{x})$ satisfying constraint \eqref{asymp_performance} is not a Lyapunov function as in \cite{ames2016control} since the function $w(\bm{x})$ itself is not required to be sign definite. 

Constraint \eqref{asymp_safety} is convex, i.e., the set of controllers satisfying constraint \eqref{asymp_safety} is convex. Therefore, we first synthesize a safe controller to ensure safety of system \eqref{control_system} via solving the following optimization:
\begin{equation}
\label{step_1}
\begin{split}
        &\bm{u}^*(\bm{x})=\arg\min_{\bm{u}(\bm{x})\in \mathcal{U}} \|\bm{u}(\bm{x})-\bm{k}(\bm{x})\|\\
s.t.&~\mathcal{L}_{h,\bm{u}}(\bm{x})\geq 0, \forall \bm{x}\in \overline{\mathcal{C}\setminus \mathcal{X}_r}.
        \end{split}
\end{equation}

After computing a safe controller $\bm{u}^*(\bm{x})$,  we then solve the following optimization:
\begin{equation}
\label{step_2}
\begin{split}
        &(w^*(\bm{x}),\delta^*)=\arg\min_{w(\bm{x}),\delta} \delta\\
s.t.~&\mathcal{L}_{w,\bm{u}^*}(\bm{x})-h(\bm{x})\geq -\delta, \forall \bm{x}\in \overline{\mathcal{C}\setminus \mathcal{X}_r};\\
&\delta\geq 0.
        \end{split}
\end{equation}

If $\delta^*=0$, the controller $\bm{u}^*(\bm{x})$ is a reach-avoid controller  with respect to the safe set $\mathcal{C}$ and target set $\mathcal{X}_r$.  Otherwise,  the controller $\bm{u}^{*}(\bm{x})$ degrades the reachability proformance. 
\begin{proposition}
\label{deltada0}
Suppose $\delta^*\in [0,\infty)$ is obtained by solving optimization \eqref{step_2}. If the set $\mathcal{D}=\{\bm{x}\in \mathbb{R}^n\mid h(\bm{x})>\delta^*\}$ has nonempty intersection with the target set $\mathcal{X}_r$, the controller $\bm{u}^{*}(\bm{x})$ is a reach-avoid controller with respect to $\mathcal{D}$ and $\mathcal{X}_r$.
\end{proposition}

\begin{remark}
\label{zeroing_relax}
Like zeroing control barrier functions in \cite{ames2016control}, a relaxation of the first condition, i.e., $\mathcal{L}_{h,\bm{u}}(\bm{x})\geq 0, \forall \bm{x}\in \overline{\mathcal{C}\setminus \mathcal{X}_r}$, in constraint  \eqref{control_g_b_f} is 
\begin{equation*}
\begin{cases}
&\mathcal{L}_{h,\bm{u}}(\bm{x})\geq -\beta h(\bm{x}), \forall \bm{x}\in \overline{\mathcal{C}\setminus \mathcal{X}_r},\\
&\beta>0.
\end{cases}
\end{equation*}
In such circumstances, we have to intensify the second condition, i.e., $\mathcal{L}_{w,\bm{u}}(\bm{x})-h(\bm{x})\geq 0, \forall \bm{x}\in \overline{\mathcal{C}\setminus \mathcal{X}_r}$, in constraint \eqref{control_g_b_f} using \[\mathcal{L}_{w,\bm{u}}(\bm{x})-h(\bm{x})>0, \forall \bm{x}\in \overline{\mathcal{C}\setminus \mathcal{X}_r}.\] Consequently, a set of new constraints, which can replace the set of constraints \eqref{control_g_b_f}, is given below:
\begin{equation*}
\begin{cases}
&\mathcal{L}_{h,\bm{u}}(\bm{x})\geq -\beta h(\bm{x}), \forall \bm{x}\in \overline{\mathcal{C}\setminus \mathcal{X}_r}, \\
&\mathcal{L}_{w,\bm{u}}(\bm{x})-h(\bm{x})>0, \forall \bm{x}\in \overline{\mathcal{C}\setminus \mathcal{X}_r}, \\
&\beta>0.
\end{cases}
\end{equation*}
Further, since $h(\bm{x})\geq 0$ for $\bm{x}\in \overline{\mathcal{C}\setminus \mathcal{X}_r}$, the above constraints can be weakened by
\begin{equation}
\label{zeroing_barrer}
\begin{cases}
&\mathcal{L}_{h,\bm{u}}(\bm{x})\geq -\beta h(\bm{x}), \forall \bm{x}\in \overline{\mathcal{C}\setminus \mathcal{X}_r}, \\
&\mathcal{L}_{w,\bm{u}}(\bm{x})>0, \forall \bm{x}\in \overline{\mathcal{C}\setminus \mathcal{X}_r}, \\
&\beta>0.
\end{cases}
\end{equation}
However, if we replace  the constraint in \eqref{step_1} with 
\begin{equation*}
\begin{cases}
\mathcal{L}_{h,\bm{u}}(\bm{x})\geq -\beta h(\bm{x}), \forall \bm{x}\in \overline{\mathcal{C}\setminus \mathcal{X}_r},\\
\beta>0,
\end{cases}
\end{equation*}
 to obtain a safe controller $\bm{u}^*(\bm{x})$,  we cannot obtain a conclusion as in Proposition \ref{deltada0} via solving \eqref{step_2} with constraints $\mathcal{L}_{w,\bm{u}^*}(\bm{x})>-\delta, \forall \bm{x}\in \overline{\mathcal{C}\setminus \mathcal{X}_r}$ and $\delta\geq 0$ when $\delta^*>0$.  $\hfill\blacksquare$
\end{remark}

If $\delta^*>0$, we  proceed to obtain a new  controller $\bm{u}^{**(}\bm{x})$ and a smaller $\delta^{**}$ via solving optimization \eqref{q_P_r} with $w^*(\bm{x})$.

\begin{equation}
    \label{q_P_r}
     \begin{split}
        &(\bm{u}^{**}(\bm{x}),\delta^{**})=\arg\min_{\bm{u}(\bm{x})\in \mathcal{U},\delta} \|\bm{u}(\bm{x})-\bm{k}(\bm{x})\|+c\|\delta\|\\
        s.t.~&\mathcal{L}_{h,\bm{u}}(\bm{x})\geq 0,  \forall \bm{x}\in \overline{\mathcal{C}\setminus \mathcal{X}_r};\\
        &\mathcal{L}_{w^*,\bm{u}}(\bm{x})-h(\bm{x})\geq -\delta,  \forall \bm{x}\in \overline{\mathcal{C}\setminus \mathcal{X}_r};\\
        &\delta\geq 0.
     \end{split}
     \end{equation}
     where $c$ is a large constant that penalizes reachability violations, similar to the one in optimization \eqref{qu}. 
     
Repeating the above procedure leads to an iterative algorithm for solving optimization \eqref{qua_asym0}, which is summarized in Alg. \ref{alg_1}. When $\delta=0$, a reach-avoid controller $\bm{u}(\bm{x}) \in \mathcal{K}_a(\bm{x})$ is generated successfully. 
\begin{algorithm}
\caption{An iterative procedure for solving optimization \eqref{qua_asym0} when a reach-avoid controller is not returned via solving optimization \eqref{qu}.}
Let $\epsilon', \epsilon^*>0$  be specified thresholds.
\begin{algorithmic}
\STATE $i:=0$;
\STATE solve optimization \eqref{step_1} to obtain a safe controller $\bm{u}_i(\bm{x})$;
\WHILE{TRUE}
\STATE compute $(w_i(\bm{x}),\delta_i)$ via solving optimization \eqref{step_2} with $\bm{u}_i(\bm{x})$;
\STATE compute $(\bm{u}_{i+1}(\bm{x}),\delta_{i+1})$  via solving  \eqref{q_P_r} with $w_i(\bm{x})$;
\IF {$\delta_i=\delta_{i+1}=0$}
\IF{$\xi_{i+1}-\xi_i \leq -\epsilon'$, where $\xi_{i+1}=\|\bm{u}_{i+1}(\bm{x})-\bm{k}(\bm{x})\|$}
\STATE $i:=i+1$;
\ELSE
\STATE return $\bm{u}_{i+1}(\bm{x})$ and terminate;
\ENDIF
\ELSE
\IF{$\delta_{i+1}-\delta_i\leq -\epsilon^{*}$}
\STATE $i:=i+1$;
\ELSE
\STATE return $\bm{u}_{i+1}(\bm{x})$ and terminate;
\ENDIF
\ENDIF
\ENDWHILE
\end{algorithmic}
\label{alg_1}
\end{algorithm}

Although constraint \eqref{control_g_b_f} is weaker than  \eqref{exponential_control}, it is still strict in practical applications. The condition $\mathcal{L}_{h,\bm{u}}(\bm{x})\geq 0, \forall \bm{x}\in \overline{\mathcal{C}\setminus \mathcal{X}_r}$  not only ensures invariance of the set $\overline{\mathcal{C}\setminus \mathcal{X}_r}$, but also enforces invariance of every positive sublevel set of the asymptotic control guidance-barrier function (if trajectories do not enter the target set $\mathcal{X}_r$). This is overly conservative.  Thus, solving optimization \eqref{qua_asym0} may lead to either a failure in synthesizing a reach-avoid controller with respect to the safe set $\mathcal{C}$ and target set $\mathcal{X}_r$,  corresponding to $\delta>0$,  or a pessimistic feedback controller, i.e., $\|\bm{u}(\bm{x})-\bm{k}(\bm{x})\|$ is large.  

\subsubsection{Lax Control Guidance-barrier Functions}
\label{IMCGBF}
In this subsection we will further relax constraints in asymptotic guidance-barrier functions and obtain less restrictive ones, expanding the space of reach-avoid controllers.  The construction of new constraints is based on a tightened set $\mathcal{D}$, which is a subset of the safe set $\mathcal{C}$.

The tightened set $\mathcal{D}$ can be any subset of the set $\mathcal{C}$ satisfying the following conditions: 
\begin{enumerate}
    \item $\mathcal{D}\subseteq \mathcal{C}$;
    \item $\partial \mathcal{D}\cap \partial \mathcal{C}=\emptyset$;
    \item $\mathcal{D}\cap \mathcal{X}_r\neq \emptyset$. 
\end{enumerate}
 In this paper we take $\mathcal{D}=\{\bm{x}\in \mathbb{R}^n\mid h'(\bm{x})\geq 0\}$, where $h'(\bm{x})=h(\bm{x})-\epsilon_0$ with $\epsilon_0>0$. It is observed that $\epsilon_0$ can be an arbitrary positive value, and $\lim_{\epsilon_0\rightarrow 0^{+}}\mathcal{D}=\mathcal{C}$. 

Based on the set $\mathcal{D}$, we define a lax control guidance-barrier function.
\begin{definition}
Let $\mathcal{C}\subseteq \mathbb{R}^n$ and $\mathcal{X}_r$  satisfy Assumption \ref{assump}  and $\mathcal{D}$ be the tightened set, then $h(\cdot):\mathbb{R}^n \rightarrow \mathbb{R}$ is called a lax control guidance-barrier function, if there exist a continuously differentiable function $w(\cdot): \mathbb{R}^n\rightarrow \mathbb{R}$ and $\beta \geq 0$ such that
\begin{equation}
\label{control_g_b_f_c_new}
\begin{cases}
&\sup_{\bm{u}(\bm{x})\in \mathcal{U}} \mathcal{L}_{h,\bm{u}}(\bm{x}) \geq -\beta h(\bm{x}), \forall \bm{x} \in \overline{(\mathcal{C}\setminus \mathcal{D})\setminus \mathcal{X}_r},\\
&\sup_{\bm{u}(\bm{x})\in \mathcal{U}} \mathcal{L}_{w,\bm{u}}(\bm{x})>0, \forall \bm{x}\in \overline{\mathcal{C}\setminus \mathcal{X}_r}.
\end{cases}
\end{equation}
\end{definition}

\begin{theorem}
\label{improved}
Let $\mathcal{C}\subseteq \mathbb{R}^n$  and $\mathcal{X}_r$ satisfy Assumption \ref{assump}  and $\mathcal{D}$ be a tightened set defined above, if $h(\bm{x})$ is a lax control guidance-barrier function, then any Lipschitz continuous controller $\bm{u}(\cdot):\mathcal{C}\rightarrow \mathcal{U}$ satisfying $\bm{u}(\bm{x})\in \mathcal{K}_l(\bm{x})$  is a reach-avoid controller with respect to the safe set $\mathcal{C}$ and target set $\mathcal{X}_r$, where
\begin{equation*}
\mathcal{K}_l(\bm{x})=\left\{\bm{u}(\bm{x})\in \mathcal{U} \middle|\;
\text{constraint}~\eqref{improve_control_c}~\text{holds}
\right
\}.
\end{equation*}
where 
\begin{equation}
\label{improve_control_c}
\begin{cases}
&\mathcal{L}_{h,\bm{u}}(\bm{x}) \geq -\beta h(\bm{x}), \forall \bm{x}\in \overline{(\mathcal{C}\setminus \mathcal{D})\setminus \mathcal{X}_r },\\
&\mathcal{L}_{w,\bm{u}}(\bm{x})>0, \forall \bm{x}\in \overline{\mathcal{C}\setminus \mathcal{X}_r},\\
&\beta\geq 0.
\end{cases}
\end{equation}
\end{theorem}

 \begin{remark}
 Constraint \eqref{improve_control_c} also applies to the case that the safe set $\mathcal{C}$ is compact.  $\hfill\blacksquare$
 \end{remark}

In constraint \eqref{improve_control_c}, when $\beta=0$, the condition $
\mathcal{L}_{h,\bm{u}}(\bm{x})\geq -\beta h(\bm{x}), \forall \bm{x}\in \overline{(\mathcal{C}\setminus \mathcal{D})\setminus \mathcal{X}_r}$ ensures invariance of the set $\overline{\mathcal{D}'\setminus \mathcal{X}_r}$ until system \eqref{control_system} enters the target set $\mathcal{X}_r$, where $\mathcal{D}'=\{\bm{x}\in \mathcal{C}\mid h(\bm{x})\leq \epsilon_1\}$ with $\epsilon_1\leq \epsilon_0$. It does not ensure invariance of the set $\{\bm{x}\in \mathcal{C}\mid h(\bm{x})\geq \epsilon_1\}$ with $\epsilon_1>\epsilon_0$, and thus it is weaker than the condition $\mathcal{L}_{h,\bm{u}}(\bm{x})\geq 0, \forall \bm{x}\in \overline{\mathcal{C}\setminus \mathcal{X}_r}$
in constraint \eqref{control_g_b_f}. However, when $\beta>0$, we can just ensure invariance of the set $\mathcal{C}\setminus \mathcal{X}_r$ until system \eqref{control_system} enters the target set $\mathcal{X}_r$. However, since $h(\bm{x})\geq 0$ for $\bm{x}\in \overline{\mathcal{C}\setminus \mathcal{X}_r}$, we cannot conclude that if $\mathcal{L}_{w,\bm{u}}(\bm{x})\geq h(\bm{x}), \forall \bm{x}\in \overline{ \mathcal{C}\setminus \mathcal{X}_r}$ holds, $\mathcal{L}_{w,\bm{u}}(\bm{x})>0, \forall \bm{x}\in \overline{\mathcal{C}\setminus \mathcal{X}_r}$ holds. It is worth remarking here that constraint \eqref{improve_control_c} is weaker than \eqref{zeroing_barrer}. Furthermore, compared to exponential and asymptotic control guidance-barrier functions, a reach-avoid controller generated by a lax one does not require $h(\bm{\phi}_{\bm{x}_0}(t))$ to be monotonically increasing with respect to $t$ before the trajectory $\bm{\phi}_{\bm{x}_0}(t)$ enters the target set $\mathcal{X}_r$, thereby diversifying the application scenarios further.

\begin{example}
Consider  a simple system:
\begin{equation}
\begin{cases}
&\dot{x}=-x+0.3,\\
&\dot{y}=-y+u_1,
\end{cases}
\end{equation}
where $\mathcal{C}=\{(x,y)^{\top}\mid 1-x^2-y^2>0\}$, $\mathcal{X}_r=\{(x,y)^{\top}\mid (x-0.3)^2+y^2-0.01 < 0\}$, and $\mathcal{U}=\{u_1\mid -0.1\leq u_1\leq 0.1\}$.

Since $h(0,0)=1$ and $h(x,y)=1-x^2-y^2<1$  for $(x,y)^{\top}\in \mathcal{X}_r$, we cannot obtain a reach-avoid controller with respect to the safe set $\mathcal{C}$ and target set $\mathcal{X}_r$ with exponential and asymptotic guidance-barrier functions. However, we can obtain a reach-avoid controller $u_1=0$ with lax guidance-barrier functions.   $h(x,y)$ is a lax guidance-barrier function satisfying \eqref{improve_control_c} with $\beta=2$, $w(x,y)=-(x-0.3)^2-y^2$, $\mathcal{D}=\{(x,y)^{\top}\mid 0.99-x^2-y^2\geq 0\}$, and $u_1=0$.
$\hfill\blacksquare$
\end{example}

 According to Proposition \ref{improved}, Problem \ref{problem1} can be transformed into the following optimization \eqref{q_P_i}
\begin{equation} 
\label{q_P_i}
\begin{split}
        &\min_{\bm{u}(\bm{x})\in \mathcal{U},w(\bm{x},)\delta,\beta} \|\bm{u}(\bm{x})-\bm{k}(\bm{x})\| +c\delta\\
s.t.~&\mathcal{L}_{h,\bm{u}}(\bm{x})\geq -\beta h(\bm{x}), \forall \bm{x}\in \overline{(\mathcal{C}\setminus \mathcal{D})\setminus\mathcal{X}_r};\\
        &\mathcal{L}_{w,\bm{u}}(\bm{x})>-\delta, \forall \bm{x}\in \overline{ \mathcal{C}\setminus \mathcal{X}_r};\\
        &\delta\geq 0,\\
        &\beta\geq 0,
        \end{split}
\end{equation}
where $\bm{u}(\bm{x}): \overline{\mathcal{C}}\rightarrow \mathcal{U}$ is a locally Lipschitz parameterized function with some unknown parameters, and $c$ is a large constant that penalizes reachability violations.

Similar to optimization \eqref{qua_asym0}, optimization \eqref{q_P_i} is also nonlinear. Thus, we also decompose it into two convex sub-problems and solve it using an iterative algorithm, which is described in Alg. \ref{alg_2}.

\begin{algorithm}[h]
\caption{An iterative procedure for solving optimization \eqref{q_P_i}.}
Let $\epsilon', \epsilon^*>0$ be specified thresholds.
\begin{algorithmic}
\STATE $i:=0$;
\IF{a Lipschitz controller $\bm{u}(\bm{x})$ is  computed via solving optimization  \eqref{qua_asym0}}
\STATE $\bm{u}_0(\bm{x}):=\bm{u}(\bm{x})$;
\ELSE
\STATE compute a Lipschitz safe controller $\bm{u}_0(\bm{x})$ satisfying 
    \begin{equation}
    \label{im_0}
        \begin{split}
           &\min_{\bm{u}(\bm{x})\in \mathcal{U}}\|\bm{u}(\bm{x})-\bm{k}(\bm{x})\| \\
        \text{s.t.~}&\mathcal{L}_{h,\bm{u}}(\bm{x})\geq 0, \forall \bm{x}\in \overline{(\mathcal{C}\setminus \mathcal{D})\setminus\mathcal{X}_r}.
    \end{split}
      \end{equation}
\ENDIF
\WHILE{TRUE}
\STATE compute $(w_i(\bm{x}),\delta_i)$ satisfying 
    \begin{equation}
    \label{im_1}
        \begin{split}
           &\min_{w(\bm{x}),\delta}\delta \\
        \text{s.t.~}&\mathcal{L}_{w,\bm{u}_i}(\bm{x})> -\delta, \forall \bm{x}\in \overline{\mathcal{C}\setminus \mathcal{X}_r};\\
        &\delta\geq 0.
    \end{split}
      \end{equation}
\STATE solve the following program to obtain $(\bm{u}_{i+1}(\bm{x}),\delta_{i+1}))$:    
    \begin{equation} 
    \label{im_2}
\begin{split}
        &\min_{\bm{u}(\bm{x})\in \mathcal{U},\delta,\beta} \|\bm{u}(\bm{x})-\bm{k}(\bm{x})\|+c\delta\\
s.t.~ &\mathcal{L}_{h,\bm{u}}(\bm{x})\geq -\beta h(\bm{x}), \forall \bm{x}\in \overline{(\mathcal{C}\setminus \mathcal{D})\setminus\mathcal{X}_r},\\
        &\mathcal{L}_{w_i^*,\bm{u}}(\bm{x})> -\delta, \forall \bm{x}\in \overline{\mathcal{C}\setminus \mathcal{X}_r},\\
        &\delta\geq 0,\\
        &\beta\geq 0.
        \end{split}
\end{equation}
\IF {$\delta_i=\delta_{i+1}=0$}
\IF{$\xi_{i+1}-\xi_i\leq -\epsilon'$, where $\xi_{i+1}=\|\bm{u}_{i+1}(\bm{x})-\bm{k}(\bm{x})\|$}
\STATE $i:=i+1$;
\ELSE
\STATE return $\bm{u}_{i+1}(\bm{x})$ and terminate;
\ENDIF
\ELSE
\IF{$\delta_{i+1}-\delta_i\leq -\epsilon^{*}$}
\STATE $i:=i+1$;
\ELSE
\STATE return $\bm{u}_{i+1}(\bm{x})$ and terminate;
\ENDIF
\ENDIF
\ENDWHILE
\end{algorithmic}
\label{alg_2}
\end{algorithm}

\begin{remark}
Suppose that $\delta_i$($i=1,\ldots$) is computed by solving optimization \eqref{im_1} or \eqref{im_2}.  If $\delta_i=0$, then the Lipschitz controller $\bm{u}_i(\bm{x})$ is a controller satisfying Theorem \ref{improved}.   
Otherwise, we can neither ensure that the controller $\bm{u}_i(\bm{x})$ will drive system  \eqref{control_system} starting from $\mathcal{C}$ to enter the target set $\mathcal{X}_r$ eventually nor ensure that the controller $\bm{u}_i(\bm{x})$ will drive system  \eqref{control_system} starting from $\{\bm{x}\in \mathbb{R}^n\mid h(\bm{x})>\delta_i\}$ to enter $\mathcal{X}_r$ eventually.
\end{remark}

\begin{remark}
It is observed that the smaller $\epsilon_0$ in defining $\mathcal{D}$ is, the weaker constraint \eqref{improve_control_c} is. However, $\epsilon_0$ cannot be zero. Thus, in numerical computations, we do not recommend the use of too small $\epsilon_0$ due to numerical errors.
\end{remark}

%% file: stochastic.tex
\section{Reach-avoid Controllers Synthesis for Stochastic Systems}
\label{RCSSS}

This section focuses on synthesizing reach-avoid controllers for stochastic systems modelled by stochastic differential equations. 

\subsection{Preliminaries}
\label{RCSSS_Pre}
In this subsection we introduce  stochastic systems and reach-avoid controllers synthesis problems of interest.

Consider an affine stochastic control system, 
\begin{equation}
    \label{SDE}
    \begin{split}
d\bm{x}(t,\bm{w})=(\bm{f}(\bm{x}(t,\bm{w}))+&\bm{g}(\bm{x}(t,\bm{w}))\bm{u}(\bm{x}(t)))dt\\
&+\bm{\sigma}(\bm{x}(t,\bm{w})) d\bm{W}(t,\bm{w}),
\end{split}
\end{equation}
where $\bm{f}(\cdot):\mathbb{R}^n\rightarrow \mathbb{R}^n$, $\bm{g}(\cdot): \mathbb{R}^n \rightarrow \mathbb{R}^{n\times m}$,  and $\bm{\sigma}(\cdot): \mathbb{R}^n \rightarrow \mathbb{R}^{n\times k}$ are locally Lipschitz continuous function; $\bm{u}(\cdot): \mathbb{R}^n \rightarrow \mathcal{U}$ with $\mathcal{U} \subset \mathbb{R}^m$ is the admissible input; $\bm{W}(t,\bm{w}): \mathbb{R}\times \Omega\rightarrow \mathbb{R}^k$ is an $k$-dimensional Wiener process (standard Brownian motion), and $\Omega$, equipped with the probability measure $\mathbb{P}$,  is the sample space $\bm{w}$ belongs to.  The expectation with respect to $\mathbb{P}$ is denoted by $\mathbb{E}[\cdot]$.

Assume that each component of $\bm{f}(\bm{x})$, $\bm{g}(\bm{x})$ and $\bm{\sigma}(\bm{x})$ is locally Lipschitz continuous. Given a locally Lipschitz controller $\bm{u}(\bm{x})$, then for an initial state $\bm{x}_0\in \mathbb{R}^n$, a stochastic differential equation \eqref{SDE} has a unique (maximal local) strong solution over some time interval $[0,T^{\bm{x}_0}(\bm{w}))$ for $\bm{w}\in \Omega$, where $T^{\bm{x}_0}(\bm{w})$ is a positive real value or infinity. We denote it as $\bm{\phi}_{\bm{x}_0}^{\bm{w}}(\cdot): [0,T^{\bm{x}_0}(\bm{w}))\rightarrow \mathbb{R}^n$, which satisfies the stochastic integral equation,
\begin{equation*}
\begin{split}
  \bm{\phi}_{\bm{x}_0}^{\bm{w}}(t)&=\bm{x}_0+ \int_{0}^t (\bm{f}(\bm{\phi}_{\bm{x}_0}^{\bm{w}}(\tau))+\bm{g}(\bm{\phi}_{\bm{x}_0}^{\bm{w}}(\tau))\bm{u}(\bm{\phi}_{\bm{x}_0}^{\bm{w}}(\tau)))d \tau\\
   &+\int_{0}^t \bm{\sigma}(\bm{\phi}_{\bm{x}_0}^{\bm{w}}(\tau)) d\bm{W}(\tau,\bm{w}).
   \end{split}
\end{equation*}

The infinitesimal generator underlying system \eqref{SDE} is presented in Definition \ref{inf_generator}.
\begin{definition}
\label{inf_generator}
Given system \eqref{SDE}  with a locally Lipschitz controller $\bm{u}(\bm{x})$,  the infinitesimal generator of a twice continuously differentiable function  $v(\bm{x})$ is defined by 
\begin{equation*}
    \begin{split}
        &\mathcal{L}_{v,\bm{u}}(\bm{x}_0)=\lim_{t\rightarrow 0}\frac{\mathbb{E}[v(\bm{\phi}_{\bm{x}_0}^{\bm{w}}(t))]-v(\bm{x}_0)}{t}\\
        &=[\frac{\partial v}{\partial \bm{x}}(\bm{f}(\bm{x})+\bm{g}(\bm{x})\bm{u}(\bm{x}))+\frac{1}{2}\textbf{tr}(\bm{\sigma}(\bm{x})^{\top}\frac{\partial^2 v}{\partial \bm{x}^2} \bm{\sigma}(\bm{x}))]\mid_{\bm{x}=\bm{x}_0}.
    \end{split}
\end{equation*}
\end{definition}

As a stochastic generalization of the Newton-Leibniz axiom, Dynkin's formula gives the expected value of any suitably smooth function of an It\^o diffusion at a stopping time.
\begin{theorem}[Dynkin's formula, \cite{oksendal2003}]
\label{dyn_theom}
Given system \eqref{SDE}  with a locally Lipschitz controller $\bm{u}(\bm{x})$.  Suppose $\tau$ is a stopping time with $E[\tau]<\infty$, and $v\in \mathcal{C}^{2}(\mathbb{R}^n)$ with compact support. Then  
\begin{equation}
\label{dyn}
\mathbb{E}[v(\bm{\phi}_{\bm{x}}^{\bm{w}}(\tau))]=v(\bm{x})+\mathbb{E}[\int_{0}^{\tau} \mathcal{L}_{v,\bm{u}}(\bm{\phi}_{\bm{x}}^{\bm{w}}(s))ds].
\end{equation}
\end{theorem}

In Theorem \ref{dyn_theom}, if we consider a twice continuously differentiable function $f$ defined on a bounded set  $B\subseteq \mathbb{R}^n$, i.e., $v(\bm{x}) \in \mathcal{C}^2(B)$, $v$ can be any twice continuously differentiable function $v(\bm{x}) \in  \mathcal{C}^2(B)$ without the assumption of compact support. In this case, the support of $v(\bm{x})$ is of course, compact, since the support of $v(\bm{x})$ is always closed and bounded. 

Given a bounded and open safe set 
\[\mathcal{C}=\{\bm{x}\in \mathbb{R}^n\mid h(\bm{x})>0\}\] with $\partial \mathcal{C}=\{\bm{x}\in \mathbb{R}^n\mid h(\bm{x})=0\}$, and a target set 
$\mathcal{X}_r$  satisfying $\mathcal{X}_r\subseteq \mathcal{C}$ and \[h(\bm{x})\leq 1, \forall \bm{x}\in \mathcal{X}_r,\] a 
 reach-avoid controller is formulated in Definition \ref{RAC_s}.

\begin{definition}[Reach-avoid Controllers]
\label{RAC_s}
Given a locally Lipschitz continuous feedback controller $\bm{u}(\cdot):\mathbb{R}^n\rightarrow \mathcal{U}$, the reach-avoid property with respect to the safe set $\mathcal{C}$ and target set $\mathcal{X}_r$ is satisfied if, starting from any initial state $\bm{x}_0$ in $\mathcal{C}$, system \eqref{SDE} with the controller $\bm{u}(\cdot)$ will enter the target set $\mathcal{X}_r$ eventually while staying inside $\mathcal{C}$ before the first target hitting time, with a probability being larger than $h(\bm{x}_0)$, i.e., 
\[\mathbb{P}\Bigg(\Big\{\bm{w}\in \Omega \mid
\begin{aligned}
&\exists t\geq 0. \bm{\phi}_{\bm{x}_0}^{\bm{w}}(t) \in \mathcal{X}_r\bigwedge \\
&\forall \tau\in [0,t]. \bm{\phi}_{\bm{x}_0}^{\bm{w}}(\tau) \in \mathcal{C}
\end{aligned}
\Big\}
\Bigg)\geq h(\bm{x}_0).\] Correspondingly, $\bm{u}(\cdot):\mathbb{R}^n \rightarrow \mathcal{U}$ is a reach-avoid controller with respect to the safe set $\mathcal{C}$ and target set $\mathcal{X}_r$.
 \end{definition}

Similar to the case for deterministic systems in Section \ref{RACS}, we wish to synthesize a reach-avoid controller in a minimally invasive fashion. The optimization problem is formulated below.

\begin{problem}
 Suppose we are given a feedback controller $\bm{k}(\bm{x}): \mathcal{C}\rightarrow \mathbb{R}^m$ for  system \eqref{SDE},  we wish to modify this controller in a minimal way so as to guarantee satisfaction of reach-avoid specifications, i.e., solve the following optimization:
\begin{equation}
    \label{q_P}
    \begin{split}
        &\bm{u}(\bm{x})=\arg\min\|\bm{u}-\bm{k}(\bm{x})\|\\
        &\text{s.t.}~\bm{u}(\cdot): \mathbb{R}^n\rightarrow \mathcal{U} \text{~is a reach-avoid controller}\\
        &\text{~~~with respect to the safe set $\mathcal{X}$ and target set $\mathcal{X}_r$.}  
     \end{split}
\end{equation}
\end{problem}

In the sequel, the derivation of sufficient conditions for synthesizing  a reach-avoid controller $\bm{u}(\bm{x})$ comes  with a new stochastic process $\{\widehat{\bm{\phi}}_{\bm{x}_0}^{\bm{w}}(t), t\geq 0\}$ for $\bm{x}_0\in \overline{\mathcal{C}}$, which is a stopped process corresponding to $\{\bm{\phi}_{\bm{x}_0}^{\bm{w}}(t), t\in [0,T^{\bm{x}_0}(\bm{w}))\}$ and the set $\mathcal{C}\setminus \mathcal{X}_r$, i.e., \begin{equation}
\widehat{\bm{\phi}}_{\bm{x}_0}^{\bm{w}}(t)=
\begin{cases}
&\bm{\phi}_{\bm{x}_0}^{\bm{w}}(t), \text{\rm~if~}t<\tau^{\bm{x}_0}(\bm{w}),\\
&\bm{\phi}_{\bm{x}_0}^{\bm{w}}(\tau^{\bm{x}_0}(\bm{w})), \text{\rm~if~}t\geq \tau^{\bm{x}_0}(\bm{w}),
\end{cases}
\end{equation}
where \[\tau^{\bm{x}_0}(\bm{w})=\inf\{t\mid \bm{\phi}_{\bm{x}_0}^{\bm{w}}(t)\in \mathcal{X}_r \wedge \bm{\phi}_{\bm{x}_0}^{\bm{w}}(t)\in \partial \mathcal{C}\}\]
is the first time of exit of $\bm{\phi}_{\bm{x}_0}^{\bm{w}}(t)$ from the set $\mathcal{C}\setminus \mathcal{X}_r$. It is worth remarking here that if the trajectory $\bm{\phi}_{\bm{x}_0}^{\bm{w}}(t)$ escapes to infinity in finite time, it must touch the boundary of the bounded safe set $\mathcal{C}$ and thus $\tau^{\bm{x}_0}(\bm{w}) \leq T^{\bm{x}_0}(\bm{w})$. The stopped process $\widehat{\bm{\phi}}_{\bm{x}_0}^{\bm{w}}(t)$ inherits the right continuity and strong Markovian property of $\bm{\phi}_{\bm{x}_0}^{\bm{w}}(t)$. Moreover, the infinitesimal generator corresponding to $\widehat{\bm{\phi}}_{\bm{x}_0}^{\bm{w}}(t)$ is identical to the one corresponding to $\bm{\phi}_{\bm{x}_0}^{\bm{w}}(t)$ on the set $\mathcal{C}\setminus \mathcal{X}_r$, and is equal to zero outside the set $\mathcal{C}\setminus \mathcal{X}_r$. That is, 
\[\mathcal{L}_{h,\bm{u}}(\bm{x})=    \frac{\partial h}{\partial \bm{x}}(\bm{f}(\bm{x})+\bm{g}(\bm{x})\bm{u}(\bm{x}))+\frac{1}{2}\textbf{tr}(\bm{\sigma}(\bm{x})^{\top}\frac{\partial^2 h}{\partial \bm{x}^2} \bm{\sigma}(\bm{x}))\]
for $\bm{x}\in \mathcal{C}\setminus \mathcal{X}_r$, and 
\[\mathcal{L}_{h,\bm{u}}(\bm{x})=0\]
for $\bm{x}\in \partial \mathcal{C}\cup \mathcal{X}_r$ \cite{prajna2007framework,xue2022_Stochastic}. This will be implicitly assumed throughout this paper.

\subsection{Reach-avoid Controllers Synthesis}
In this section, we extend our methods for deterministic systems in Subsection \ref{egbf} and \ref{ACGBF} to stochastic systems modelled by stochastic differential equations \eqref{SDE}. The algorithms of solving the resulting optimizations are the same with the ones in Subsection \ref{egbf} and \ref{ACGBF}  and therefore, are omitted here. 

\subsubsection{Exponential Control Guidance-barrier Functions}
 In this subsection we introduce the notion of exponential control guidance-barrier functions for synthesizing reach-avoid controllers.
\begin{definition}
Let $\mathcal{C}$ and $\mathcal{X}_r$ be the sets defined in Subsection  \ref{RCSSS_Pre}, then $h(\cdot): \mathbb{R}^n \rightarrow \mathbb{R}$ is called an exponential control guidance-barrier function, if there exists $\alpha \in (0,\infty)$ such that
\begin{equation}
\sup_{\bm{u}(\bm{x})\in \mathcal{U}} \mathcal{L}_{h,\bm{u}}(\bm{x})\geq \alpha h(\bm{x}), \forall \bm{x}\in \overline{\mathcal{C}\setminus \mathcal{X}_r},\\
\end{equation}
holds.
\end{definition}

Given an exponential control guidance-barrier function, the sufficient condition for ensuring that a controller $\bm{u}(\bm{x})$ is a reach-avoid controller with respect to the safe set $\mathcal{C}$ and target set $\mathcal{X}_r$ is formulated in Theorem \ref{exponential_s}. 

\begin{theorem}
\label{exponential_s}
Given the safe set $\mathcal{C}$, target set $\mathcal{X}_r$ defined  in Subsection \ref{RCSSS_Pre},  if $h(\cdot): \mathbb{R}^n \rightarrow \mathbb{R}$ is an exponential control guidance-barrier function, then any  locally Lipschitz controller $\bm{u}(\cdot): \overline{\mathcal{C}}\rightarrow \mathcal{U}$ satisfying $\bm{u}(\bm{x}) \in \mathcal{K}_{s,e}(\bm{x})$ is a reach-avoid controller with respect to the safe set $\mathcal{C}$ and target set $\mathcal{X}_r$, where
\[\mathcal{K}_{s,e}(\bm{x})=\{\bm{u}(\bm{x})\in \mathcal{U}\mid \text{constraint}~\eqref{stochastic_c_e}~\text{holds}\}\]
with 
\begin{equation}
\label{stochastic_c_e}
\begin{cases}
&\mathcal{L}_{h,\bm{u}}(\bm{x})\geq  \alpha h(\bm{x}),  \forall \bm{x}\in \overline{\mathcal{C}\setminus \mathcal{X}_r},\\
&\alpha>0.
\end{cases}
\end{equation}
\end{theorem}


Similarly, optimization \eqref{q_P} can be encoded into the following program,
\begin{equation} 
\label{s_qu}
\begin{split}
        &\min_{\bm{u}(\bm{x})\in \mathcal{U},\alpha,\delta} \|\bm{u}(\bm{x})-\bm{k}(\bm{x})\|+c\delta\\
s.t.~&\mathcal{L}_{h,\bm{u}}(\bm{x})-\alpha h(\bm{x})\geq -\delta, \forall \bm{x}\in \overline{\mathcal{C}\setminus \mathcal{X}_r},\\
&0\leq \delta<\alpha,\\
&\alpha\geq \xi_0.
        \end{split}
\end{equation}
where $\bm{u}(\bm{x}): \overline{\mathcal{C}}\rightarrow \mathcal{U}$ is a locally Lipschitz parameterized function with some unknown parameters, $\xi_0$ is a user-defined positive value which is to enforce the strict positivity of $\alpha$, and $c$ is a large constant that penalizes safety/reachability violations. 
The set $\mathcal{K}_{s,e}(\bm{x})$ is convex, thus a direct computation of a feedback controller $\bm{u}(\bm{x})$ satisfying constraints in optimization  \eqref{s_qu} using convex optimization is possible. 

\begin{proposition}
\label{s_e}
        Suppose $(\bm{u}_0(\bm{x}),\alpha,\delta)$ is obtained by solving optimization \eqref{s_qu}. 
       \begin{enumerate}
           \item If $\delta=0$, $\bm{u}_0(\bm{x})$ is a reach-avoid controller with respect to the safe set $\mathcal{C}$ and target set $\mathcal{X}_r$.
          \item If $\delta\neq 0$ , $\bm{u}_0(\bm{x})$ is a reach-avoid controller with respect to the set $\mathcal{D}$ and target set $\mathcal{X}_r$, where $\mathcal{D}=\{\bm{x}\mid h'(\bm{x})>0\}$ with $h'(\bm{x})=\frac{\alpha h(\bm{x})-\delta}{\alpha-\delta}$.
      \end{enumerate}
\end{proposition}

\subsubsection{Asymptotic Control Guidance-barrier Functions}
In this subsection we introduce asymptotic control guidance-barrier functions for synthesizing reach-avoid controllers.

\begin{definition}
Given sets $\mathcal{C}$ and $\mathcal{X}_r$ defined in Subsection \ref{RCSSS_Pre},  then $h(\cdot): \mathbb{R}^n \rightarrow \mathbb{R}$ is called an asymptotic control guidance-barrier function, if there exists a twice continuously differentiable function $w(\cdot): \mathbb{R}^n \rightarrow \mathbb{R}$ such that 
\begin{equation}
\begin{cases}
&\sup_{\bm{u}(\bm{x})\in \mathcal{U}} \mathcal{L}_{h,\bm{u}}(\bm{x})\geq 0, \forall \bm{x}\in \overline{\mathcal{C}\setminus \mathcal{X}_r},\\
&\sup_{\bm{u}(\bm{x})\in \mathcal{U}} \mathcal{L}_{w,\bm{u}}(\bm{x})\geq h(\bm{x}), \forall \bm{x}\in \overline{\mathcal{C}\setminus \mathcal{X}_r}.\\
\end{cases}
\end{equation}
\end{definition}

The asymptotic control guidance-barrier function facilitates the construction of constraints for synthesizing reach-avoid controllers.

\begin{theorem}
\label{asymp_s}
Given sets $\mathcal{C}$  and $\mathcal{X}_r$ defined in Subsection \ref{RCSSS}, if $h(\cdot): \mathbb{R}^n \rightarrow \mathbb{R}$ is an asymptotic control guidance-barrier function, then any Lipschitz controller $\bm{u}(\cdot): \mathcal{C}\rightarrow \mathcal{U}$ satisfying $\bm{u}(\bm{x}) \in \mathcal{K}_{s,a}(\bm{x})$ is a reach-avoid controller with respect to the safe set $\mathcal{C}$ and target set $\mathcal{X}_r$, 
where 
\[\mathcal{K}_{s,a}(\bm{x})=\{\bm{u}(\bm{x})\in \mathcal{U}\mid \text{constraint}~\eqref{stochastic_c_e_a}~\text{holds}\}\]
with 
\begin{equation}
\label{stochastic_c_e_a}
\begin{cases}
&\mathcal{L}_{h,\bm{u}}(\bm{x})\geq 0, \forall \bm{x}\in \overline{\mathcal{C}\setminus \mathcal{X}_r},\\
&\mathcal{L}_{w,\bm{u}}(\bm{x})\geq h(\bm{x}), \forall \bm{x}\in \overline{\mathcal{C}\setminus \mathcal{X}_r}.\\
\end{cases}
\end{equation}
\end{theorem}

According to Theorem \ref{asymp_s}, optimization \eqref{q_P} can be encoded into the following program,
   \begin{equation} 
    \label{s_im_2}
\begin{split}
        &\min_{\bm{u}(\bm{x})\in \mathcal{U},w(\bm{x}),\delta} \|\bm{u}(\bm{x})-\bm{k}(\bm{x})\|+c\delta\\
s.t.~ &\mathcal{L}_{h,\bm{u}}(\bm{x})\geq 0, \forall \bm{x}\in \overline{\mathcal{C}\setminus\mathcal{X}_r},\\
        &\mathcal{L}_{w,\bm{u}}(\bm{x})\geq h(\bm{x})-\delta, \forall \bm{x}\in \overline{\mathcal{C}\setminus \mathcal{X}_r},\\
        &0\leq \delta<1,
        \end{split}
\end{equation}
where $\bm{u}(\bm{x}): \overline{\mathcal{C}}\rightarrow \mathcal{U}$ is a locally Lipschitz parameterized function with some unknown parameters, and $c$ is a large constant that penalizes safety/reachability violations.

\begin{proposition}
\label{s_im_2_pro}
Suppose  $(\bm{u}^*(\bm{x}),\delta^*)$ is obtained by solving optimization \eqref{s_im_2}. The controller $\bm{u}^*(\bm{x})$ is a reach-avoid controller with respect to the sets $\mathcal{D}$ and $\mathcal{X}_r$, where $\mathcal{D}=\{\bm{x}\mid h'(\bm{x})>0\}$ with $h'(\bm{x})=\frac{h(\bm{x})-\delta}{1-\delta}$.
\end{proposition}

%% file: example.tex
\section{Examples}
\label{exa}
This section demonstrates the theoretical developments of proposed methods numerically. The semi-definite programming tool Mosek \cite{mosek2015mosek} is employed to address the optimization involved. Some parameters are presented in Table \ref{tab11}, and the computed $\|\bm{u}(\bm{x})-\bm{k}(\bm{x})\|$'s are summarized in Table \ref{tab1} and \ref{tab2}. In the examples below, we attempt to synthesize a reach-avoid controller of the linear form in a minimal way of modifying the controller $\bm{k}(\bm{x})=\bm{0}$. 
\begin{table}
 \centering
 \begin{tabular}{|c|c|c|c|c|}
 \hline  \multicolumn{3}{|c|}{Optimization}\\
  \eqref{qu}/\eqref{s_qu}  & \eqref{qua_asym0}/\eqref{s_im_2}&\eqref{q_P_i}\\
 \hline     $\xi_0=10^{-3}$         &      $\epsilon'=10^{-3}$  & $\epsilon'=10^{-3}$, $\epsilon_0=10^{-3}$\\ \hline
 \end{tabular}
 \caption{Parameters in solving Optimization \eqref{qu}, \eqref{qua_asym0}, \eqref{q_P_i}, \eqref{s_qu} and \eqref{s_im_2} ($\epsilon_0$ is used to define $\mathcal{D}$). }
  \label{tab11}
 \end{table}
 
 \begin{table}
 \centering
 \begin{tabular}{|c|c|c|c|c|}
 \hline Example & \multicolumn{3}{|c|}{Optimization}\\
 &\eqref{qu} & \eqref{qua_asym0}&\eqref{q_P_i}\\
 \hline 2 &      0.2329           &         0.089  &  $3.829\times 10^{-4}$\\ \hline
          3 &      3.2002          &         3.1563  &  3.1563\\ \hline
           4 &      1.7967          &         1.7912  &  1.7372\\ \hline
 \end{tabular}
 \caption{$\|\bm{u}(\bm{x})-\bm{k}(\bm{x})\|$ for Examples 2-4. }
  \label{tab1}
 \end{table}

\begin{table}
 \centering
 \begin{tabular}{|c|c|c|c|c|}
 \hline Example & \multicolumn{2}{|c|}{Optimization}\\
 &\eqref{s_qu} & \eqref{s_im_2}\\
 \hline 5 &      93.6528          &        93.4427  \\ \hline
 \end{tabular}
 \caption{$\|\bm{u}(\bm{x})-\bm{k}(\bm{x})\|$ for Example 5. }
  \label{tab2}
 \end{table}

\subsection{Deterministic Case}
\begin{example}
\label{ex0}
Consider a simple mobile  robot navigation that can be described by the following equation \cite{romdlony2016stabilization}
\begin{equation}
\begin{cases}
\dot{x}=u_1,\\
\dot{y}=u_2,
\end{cases}
\end{equation}
where $\mathcal{C}=\{(x,y)^{\top}\mid 1-x^2-y^2>0\}$, $\mathcal{X}_r=\{(x,y)^{\top}\mid 0.01(x-0.9)^2+y^2-0.01<0\}$, and $\mathcal{U}=\{(u_1,u_2)^{\top}\mid 10^{-3}\leq u_1\leq 1, -1\leq u_2\leq 1\}$.

 The vector fields formed by the computed controllers are visualized in Fig. \ref{fig_ex3}. It is concluded from Table \ref{tab1} that the reach-avoid controller generated by exponential guidance-barrier functions can be improved by asymptotic and lax ones, and the lax guidance-barrier function generates the best controller.

\begin{figure}
\centering
   \includegraphics[width=2.05in,height=1.2in]{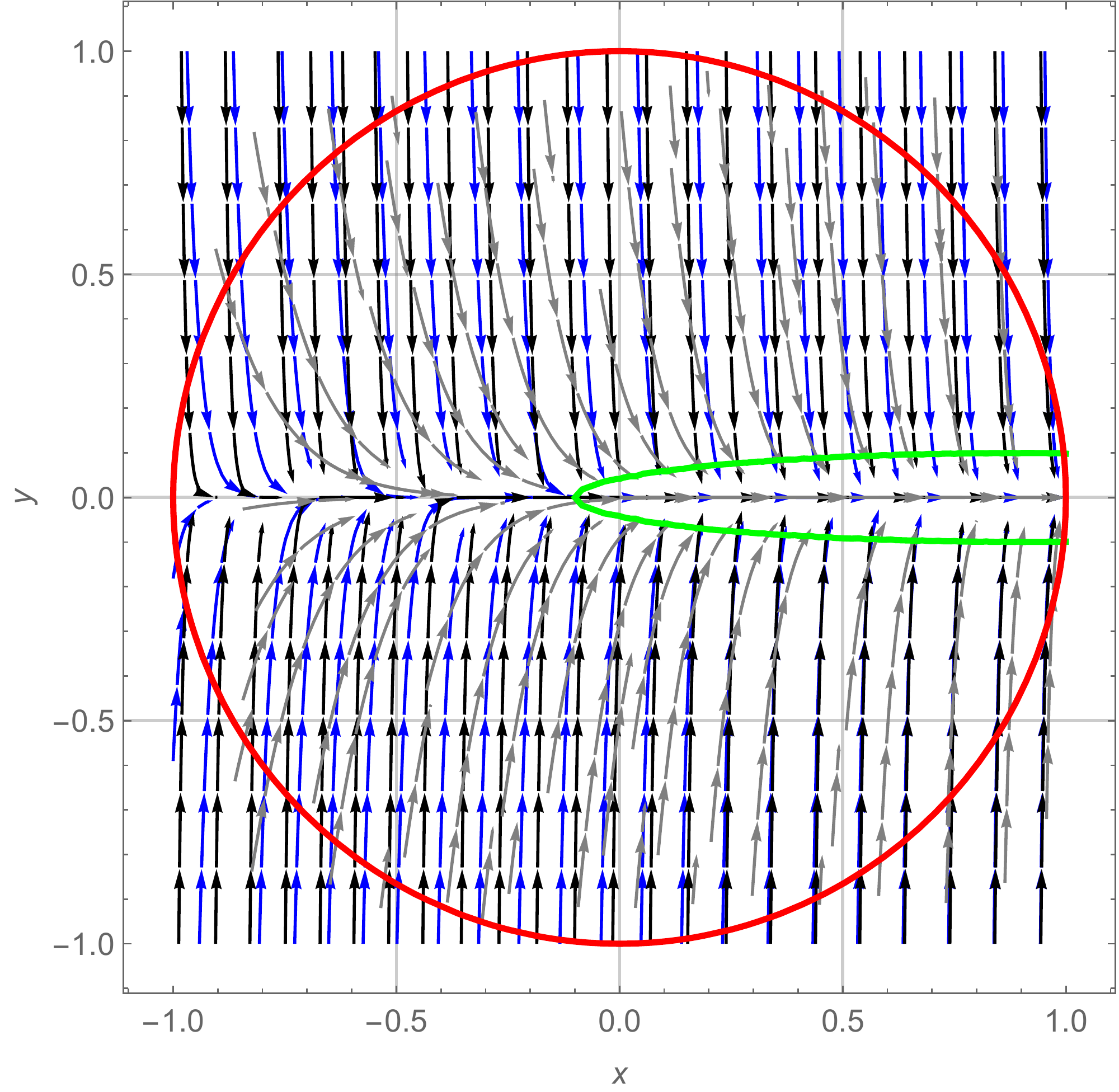}
  \caption{An illustration of controller synthesis for Example \ref{ex0}. Red and green curves denote the boundaries of the safe set $\mathcal{C}$ and target set $\mathcal{X}_r$, respectively.  Blue, black and gray curves respectively  denote the vector fields with controllers computed via solving Optimization \eqref{qu}, \eqref{qua_asym0} and \eqref{q_P_i}, respectively.}
  \label{fig_ex3}
  \end{figure}
\end{example}

\begin{example}
\label{ex3}
Consider a second order linear model from \cite{wang2022safety},
\begin{equation}
\begin{cases}
&\dot{x}=2x+y+u_1,\\
&\dot{y}=3x+y+u_2,
\end{cases}
\end{equation}
where $\mathcal{C}=\{(x,y)^{\top}\mid 1-x^2-y^2>0\}$, $\mathcal{X}_r=\{(x,y)^{\top}\mid x^2+y^2-0.01<0\}$, and $\mathcal{U}=\{(u_1,u_2)^{\top}\mid -3\leq u_1\leq 3,-3\leq u_2\leq 3\}$.

The vector fields formed by $\bm{k}(\bm{x})$ and the computed controllers are visualized in Fig. \ref{fig_ex31}. It is concluded from Table \ref{tab1} that the reach-avoid controller generated by exponential guidance-barrier functions can be improved by asymptotic and lax ones. However,the reach-avoid controller generated by the asymptotic guidance-barrier function cannot be improved further via  the lax one. 

\begin{figure}
\centering
\begin{minipage}{0.2\textwidth}
\includegraphics[width=1.5in,height=1.2in]{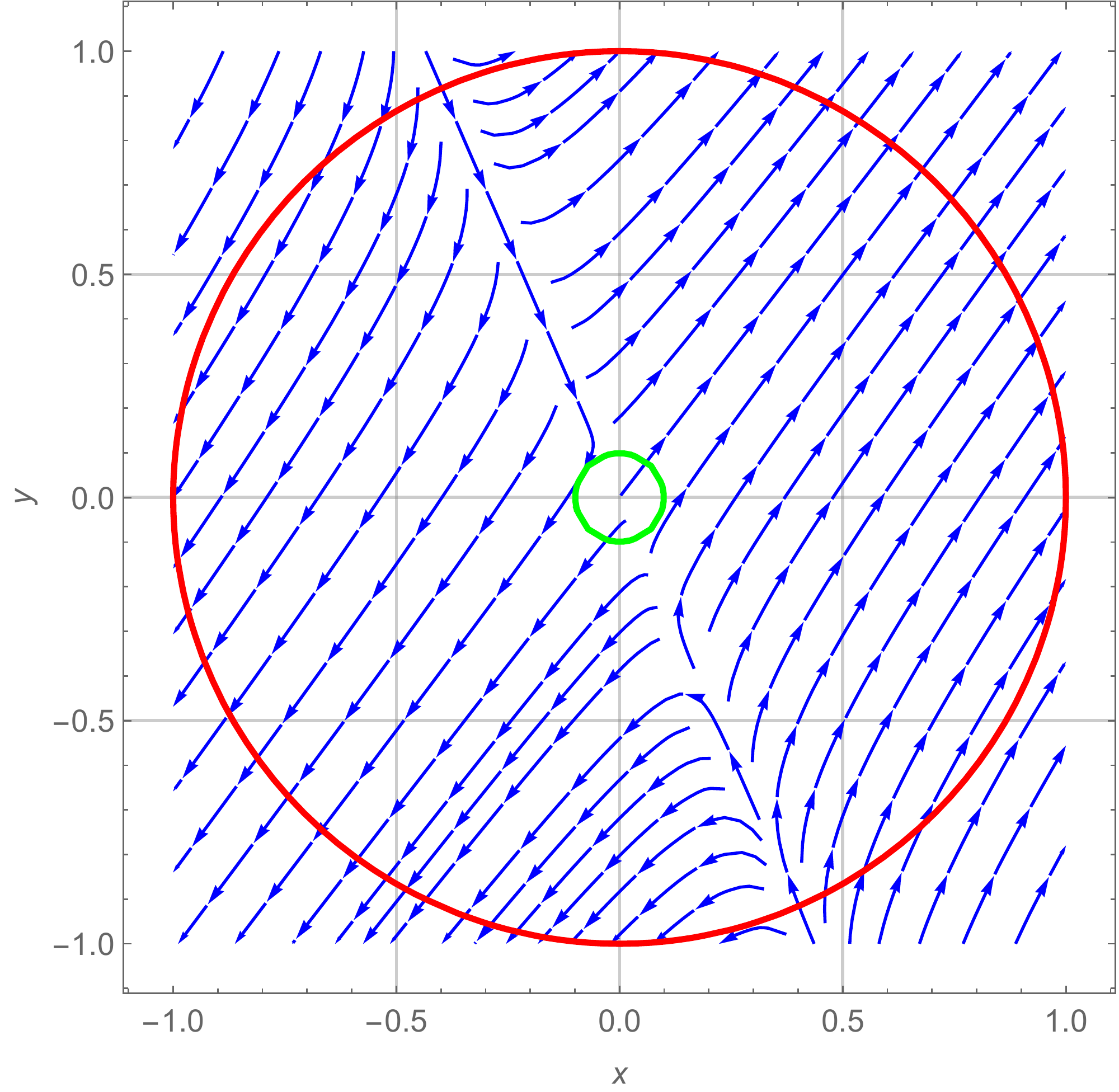}
\end{minipage}
\begin{minipage}{0.2\textwidth}
   \includegraphics[width=1.5in,height=1.2in]{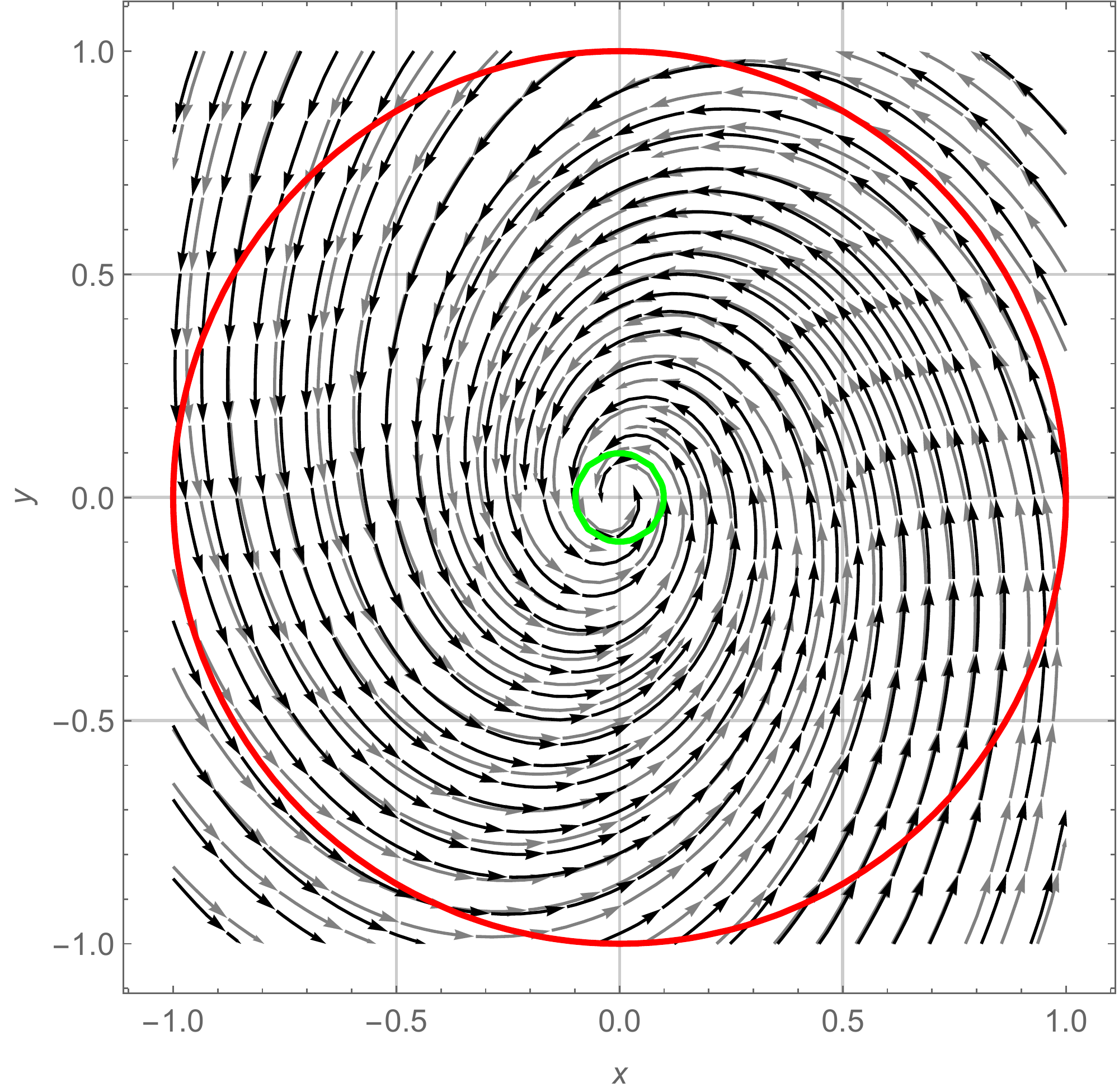}
  \end{minipage}
    \caption{An illustration of controller synthesis for Example \ref{ex3}. Red and green curves denote the boundaries of the sets $\mathcal{C}$ and $\mathcal{X}_r$, respectively. Left: the vector fields for Example \ref{ex3} with the controller $\bm{k}(\bm{x})$. Right: Gray and black curves respectively  denote the vector fields with controllers computed via solving Optimization \eqref{qu} and \eqref{qua_asym0}, respectively.} 
      \label{fig_ex31}
  \end{figure}
\end{example}

\begin{example}
\label{ex4}
    Consider a second order polynomial nonlinear control affine system from \cite{wang2022safety},
\begin{equation}
\begin{cases}
&\dot{x}=y+(x^2+y+1)u_1,\\
&\dot{y}=x+\frac{1}{3}x^3+y+(y^2+x+1)u_2,
\end{cases}
\end{equation}
where $\mathcal{C}=\{(x,y)^{\top}\mid 1-x^2-y^2>0\}$, $\mathcal{X}_r=\{(x,y)^{\top}\mid x^2+y^2-0.01<0\}$, and $\mathcal{U}=\{(u_1,u_2)^{\top}\mid -2\leq u_1\leq 2,-2\leq u_2\leq 2\}$.

 The vector fields formed by the computed controllers are visualized in Fig. \ref{fig_ex41}. It is concluded from Table \ref{tab1} that the reach-avoid controller generated by exponential guidance-barrier functions can be improved by asymptotic and lax ones, although the improvements are not significant..  

\begin{figure}
\centering
\begin{minipage}{0.2\textwidth}
\includegraphics[width=1.5in,height=1.2in]{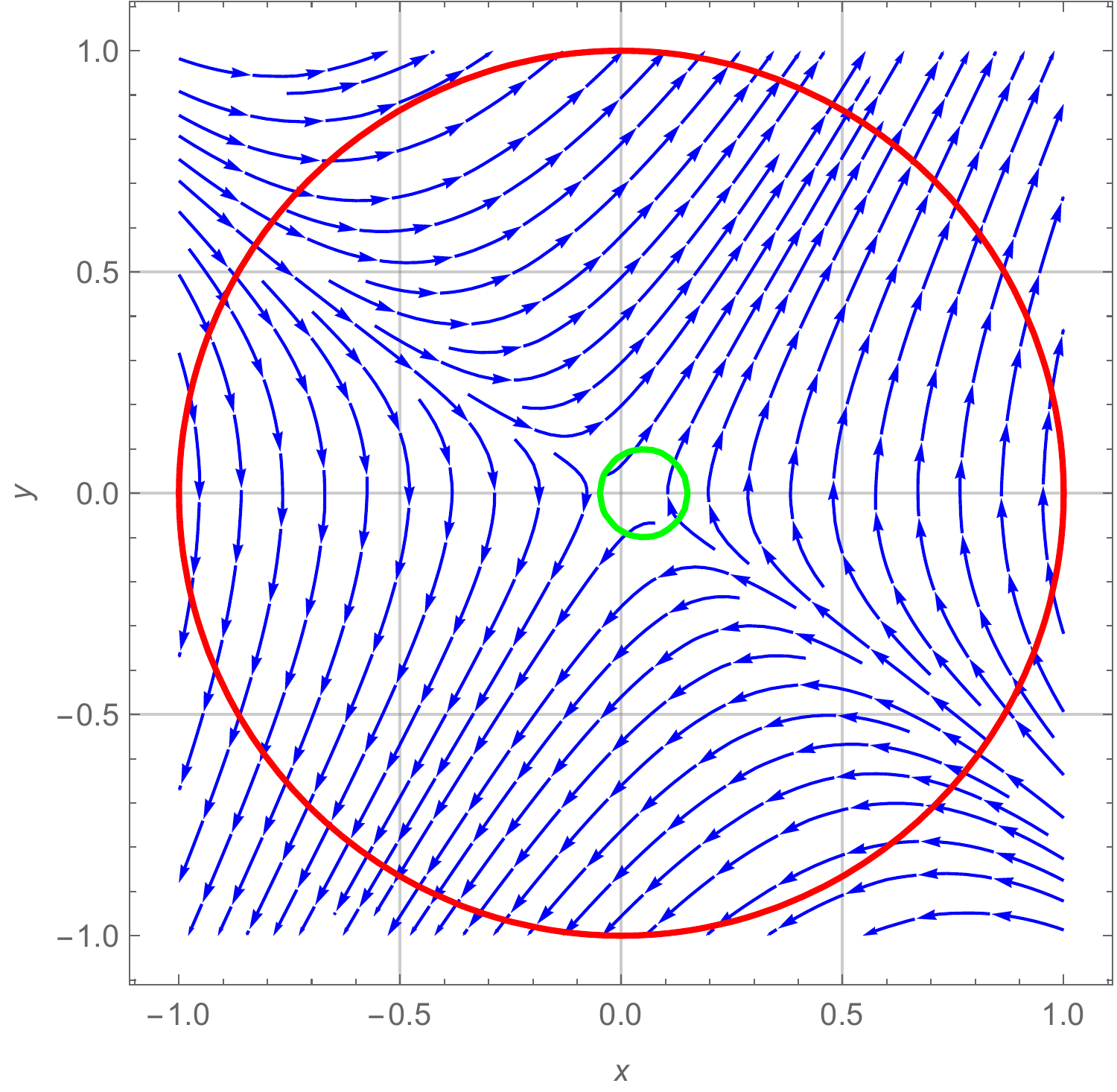}
\end{minipage}
\begin{minipage}{0.2\textwidth}
   \includegraphics[width=1.5in,height=1.2in]{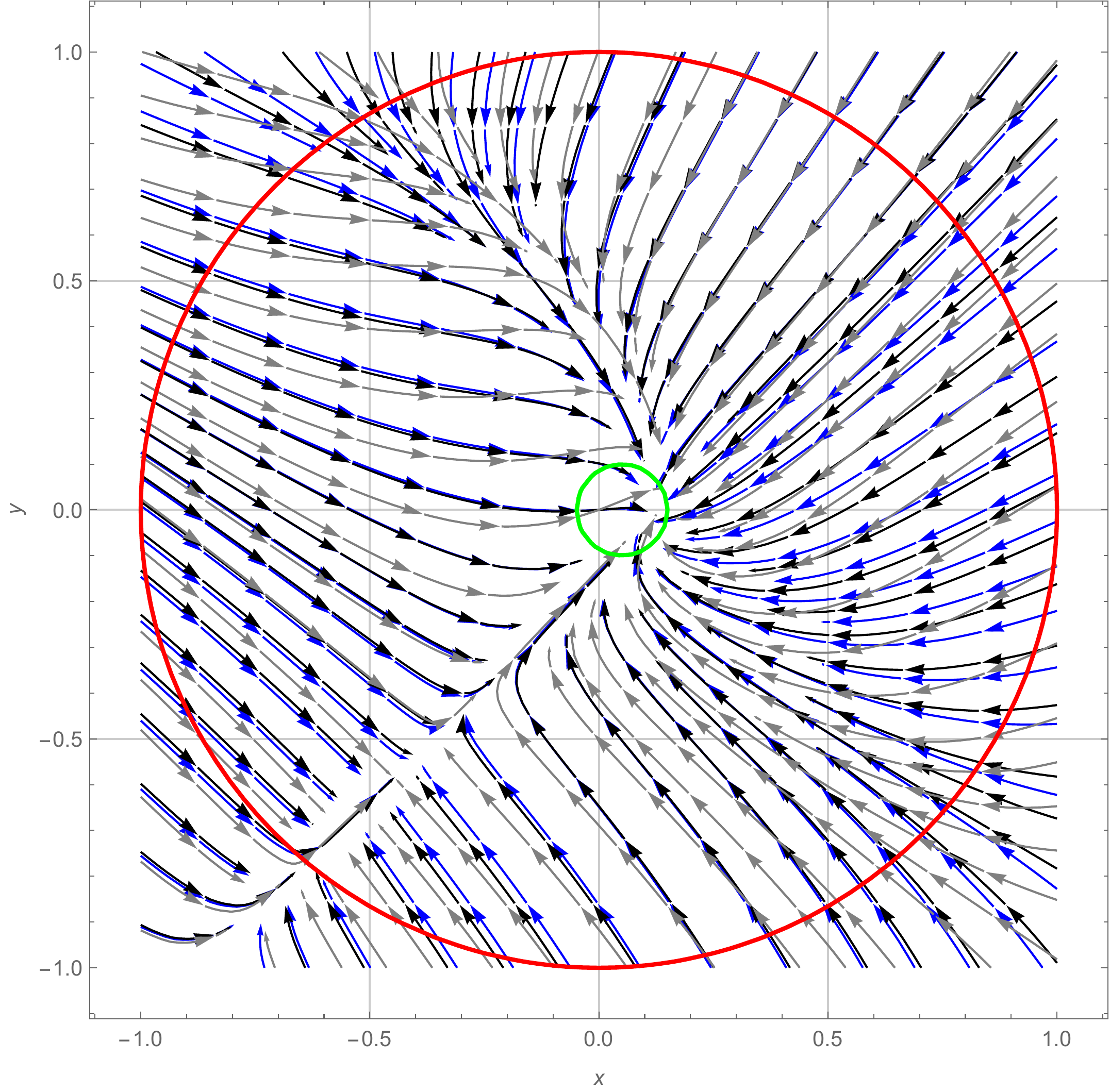}
  \end{minipage}
    \caption{An illustration of controller synthesis for Example \ref{ex4}. Red and green curves denote the boundaries of the sets $\mathcal{C}$ and $\mathcal{X}_r$, respectively. Left: the vector fields for Example \ref{ex4} with the controller $\bm{k}(\bm{x})$. Right: Blue, gray and black curves respectively  denote the vector fields with controllers computed via solving Optimization \eqref{qu} and \eqref{qua_asym0}, respectively.} 
      \label{fig_ex41}
  \end{figure}
  
\end{example}

\subsection{Stochastic Case}
\begin{example}
Consider the following stochastic system adapted from Example \ref{ex0}
\begin{equation}
\begin{cases}
\dot{x}=u_1+dw,\\
\dot{y}=u_2,
\end{cases}
\end{equation}
where $\mathcal{C}=\{(x,y)^{\top}\mid 1-x^2-y^2>0\}$, $\mathcal{X}_r=\{(x,y)^{\top}\mid 0.01(x-0.9)^2+y^2-0.01<0\}$, and $\mathcal{U}=\mathbb{R}^2$.

We synthesize a reach-avoid controller $\bm{u}=(u_1,u_2)^{\top}$. It is concluded from Table \ref{tab2} that the reach-avoid controller generated by asymptotic guidance-barrier functions improve the one from exponential guidance-barrier functions.

\end{example}

%% file: conclusion.tex
\section{Conclusion}
In this paper we investigated the problem of synthesizing reach-avoid controllers for deterministic systems modelled by ordinary differential equations and stochastic systems modelled by stochastic differential equations, based on the notion of control guidance-barrier functions. Several control guidance-barrier functions were respectively proposed for deterministic and stochastic systems to synthesizing reach-avoid controllers.  Finally, four numerical examples demonstrated the theoretical developments of proposed methodologies.  

Several factors may cause safety and reachability violations of the plant even when the Lipschitz reach-avoid controller is gained in practice. One challenge is raised by the digital/discrete implementation of a continuous-time system with continuous inputs. In a practical implementation, the system state is only observable at each sampling time, and the control signal is applied in a zero-order hold manner during each sampling period. That is, the system is implemented as a sampled-data system \cite{ghaffari2018safety}.  In future work, we would like to investigate the problem of synthesizing reach-avoid controllers for sampled-data detereminstic and stochastic systems based on control guidance-barrier functions in this paper.

%% file: appendix.tex
\section{Appendix}

\textbf{The proof of Theorem \ref{exponential_control_pro}:}

\begin{proof}
Constraint \eqref{exponential_control} implies
\[h(\bm{\phi}_{\bm{x}_0}(t))\geq e^{\lambda t}h(\bm{x}_0)>0\]
if $\bm{\phi}_{\bm{x}_0}(\tau)\in \overline{\mathcal{C}\setminus \mathcal{X}_r}$ for $\tau\in [0,t]$, where $\bm{x}_0\in \mathcal{C}$. Therefore, trajectories starting from the set $\mathcal{C}$ cannot touch the boundary  $\partial \mathcal{C}$ and thus cannot leave $\mathcal{C}$ before entering the target set $\mathcal{X}_r$.  Due to the fact that $\overline{\mathcal{C}}$ is compact and the function $h(\bm{x})$ is continuously differentiable, we have that trajectories starting from the set $\mathcal{C}$ cannot stay inside $\mathcal{C}\setminus \mathcal{X}_r$ for all the time and thus they will enter the target set $\mathcal{X}_r$ eventually. Consequently, $\bm{u}(\bm{x})$  is a reach-avoid controller with respect to the safe set $\mathcal{C}$ and target set $\mathcal{X}_r$.
\end{proof}

\textbf{The proof of Proposition \ref{deltadayu0_ex}:}
\begin{proof}
The conclusion with $\delta=0$ follows from Theorem \ref{exponential_control_pro}.

For the case that $\delta>0$, let $h'(\bm{x})=h(\bm{x})-\frac{\delta}{\lambda}$. Thus, we have that \[\mathcal{L}_{h',\bm{u}}(\bm{x})-\lambda h'(\bm{x})\geq 0, \forall \bm{x}\in \overline{\mathcal{D}\setminus \mathcal{X}_r}.\]
Thus, as argued in Theorem \ref{exponential_control_pro}, we have the conclusion.
\end{proof}

\textbf{The proof of Theorem \ref{asymp_control}:}

\begin{proof}
The constraint $\mathcal{L}_{h,\bm{u}}(\bm{x})\geq 0,  \forall \bm{x}\in \overline{\mathcal{C}\setminus \mathcal{X}_r}$ implies 
\[h(\bm{\phi}_{\bm{x}_0}(t))\geq h(\bm{x}_0)>0\]
if $\bm{\phi}_{\bm{x}_0}(\tau)\in \overline{\mathcal{C}\setminus \mathcal{X}_r}$ for $\tau\in [0,t]$, where $\bm{x}_0\in \mathcal{C}$. Therefore, trajectories starting from the set $\mathcal{C}$ cannot touch the boundary  $\partial \mathcal{C}$ and thus cannot leave  $\mathcal{C}$ before entering the target set $\mathcal{X}_r$. 

Also, \[\mathcal{L}_{w,\bm{u}}(\bm{x})-h(\bm{x})\geq 0, \forall \bm{x}\in \overline{\mathcal{C}\setminus \mathcal{X}_r}\] implies that 
\[w(\bm{\phi}_{\bm{x}_0}(t))-w(\bm{x}_0)\geq \int_{0}^{t}h(\bm{\phi}_{\bm{x}_0}(\tau))d\tau\geq h(\bm{x}_0)t\]
if $\bm{\phi}_{\bm{x}_0}(\tau)\in \overline{\mathcal{C}\setminus \mathcal{X}_r}$ for $\tau\in [0,t]$. Due to the fact that $\overline{\mathcal{C}}$ is compact and the function $w(\bm{x})$ is continuously differentiable, we have that trajectories starting from the set $\mathcal{C}$ cannot stay inside $\mathcal{C}\setminus \mathcal{X}_r$ for all the time and thus they will enter the target set $\mathcal{X}_r$ eventually. Thus,  $\bm{u}(\bm{x})$ is a reach-avoid controller with respect to the safe set $\mathcal{C}$ and target set $\mathcal{X}_r$.
\end{proof}

\textbf{The proof of Proposition \ref{deltada0}:}

\begin{proof}
Let $h'(\bm{x})=h(\bm{x})-\delta^*$. Following the arguments in Theorem \ref{asymp_control}, we have the conclusion.
\end{proof}

\textbf{The proof of Theorem \ref{improved}:}

\begin{proof}
Firstly, we show that trajectories starting from $\mathcal{C}$ cannot leave $\mathcal{C}$ if they do not enter the target set $\mathcal{X}_r$.

Assume that this is not true, that is, there exists a trajectory $\bm{\phi}_{\bm{x}_0}(t)$ starting from $\bm{x}_0\in \mathcal{C}$, which will touch the boundary $\partial \mathcal{C}$ and stay inside the set $\mathcal{C}\setminus \mathcal{X}_r$ before touching the boundary $\partial \mathcal{C}$.  Since if $\bm{x}_0\in \mathcal{D}$,  the trajectory $\bm{\phi}_{\bm{x}_0}(t)$ will enter the set $\mathcal{C}\setminus \mathcal{D}$ before touching the boundary $\partial \mathcal{C}$. Therefore, we just consider $\bm{x}_0 \in \mathcal{C}\setminus \mathcal{D}$. Assume that the first time instant of touching the boundary $\partial \mathcal{C}$ is $\tau>0$. Consequently, \[h(\bm{\phi}_{\bm{x}_0}(\tau))=0\] and  \[h(\bm{\phi}_{\bm{x}_0}(t))>0\] for $t\in [0,\tau)$. However, $h(\bm{x}_0)>0$. This contradicts the first condition, i.e., \[\mathcal{L}_{h,\bm{u}}(\bm{x}) \geq -\beta h(\bm{x}), \forall \bm{x}\in \overline{(\mathcal{C}\setminus \mathcal{D})\setminus \mathcal{X}_r },\] in constraint \eqref{improve_control_c}, which implies that $h(\bm{\phi}_{\bm{x}_0}(\tau))\geq e^{-\beta \tau} h(\bm{x}_0)>0$. Thus, trajectories starting from $\mathcal{C}$ will stay inside $\mathcal{C}$ for all the time if they do not enter the target set $\mathcal{X}_r$.

 Assume that there exists a trajectory $\bm{\phi}_{\bm{x}_0}(t)$, which stays inside $\mathcal{C}\setminus \mathcal{X}_r$ for all the time. Since
\[\mathcal{L}_{w,\bm{u}}(\bm{x})> 0, \forall \bm{x}\in \overline{\mathcal{C}\setminus \mathcal{X}_r}\]
and the set $\overline{\mathcal{C}\setminus \mathcal{X}_r}$ is compact,  we have that there exists $\delta>0$ such that 
\[\mathcal{L}_{w,\bm{u}}(\bm{x})\geq \delta, \forall \bm{x}\in \overline{\mathcal{C}\setminus \mathcal{X}_r}.\]
Consequently,
\[w(\bm{\phi}_{\bm{x}_0}(t))-w(\bm{x}_0) \geq \delta t, \forall t\in [0,+\infty)\]
and thus \[w(\bm{\phi}_{\bm{x}_0}(t))\geq t\delta +w(\bm{x}_0), \forall t\in [0,\infty),\] which contradicts that $w(\bm{x})$ is bounded over the compact set $\mathcal{C}$ since $w(\bm{x})$ is continuously differentiable. 

In summary, the conclusion holds. 
 \end{proof}

 \textbf{The proof of Theorem \ref{exponential_s}:}

 \begin{proof}
Assume that $\bm{x}_0\in \mathcal{C}$.  According to Lemma 1 in \cite{xue2022_Stochastic}, 
\begin{equation*}
\begin{split}
    &\mathbb{P}(\{\bm{w}\in \Omega\mid \exists t\geq 0. \bm{\phi}_{\bm{x}_0}^{\bm{w}}(t) \in \mathcal{X}_r\bigwedge \forall \tau\in [0,t]. \bm{\phi}_{\bm{x}_0}^{\bm{w}}(\tau) \in \mathcal{C}\})\\
    &=\mathbb{P}(\{\bm{w}\in \Omega\mid \exists t\geq 0. \widehat{\bm{\phi}}_{\bm{x}_0}^{\bm{w}}(t) \in \mathcal{X}_r\}).
    \end{split}
\end{equation*}
Consequently, we just need to show that 
\[\mathbb{P}(\{\bm{w}\in \Omega\mid \exists t\geq 0. \widehat{\bm{\phi}}_{\bm{x}_0}^{\bm{w}}(t) \in \mathcal{X}_r\})\geq h(\bm{x}_0).\]

From constraint \eqref{stochastic_c_e} and Theorem \ref{dyn_theom}, we have that 
\begin{equation}
\label{leq}
h(\bm{x}_0)\leq \mathbb{E}[h(\widehat{\bm{\phi}}_{\bm{x}_0}^{\bm{w}}(t))], \forall t\in \mathbb{R}_{\geq 0}.
\end{equation}
Also, due to the fact that $h(\bm{x})\leq 1$ for $\bm{x}\in \mathcal{X}_r$,  constraint \eqref{stochastic_c_e} indicates that 
\begin{equation*}
\begin{split}
\alpha h(\widehat{\bm{\phi}}_{\bm{x}_0}^{\bm{w}}(t))\leq \alpha 1_{\mathcal{X}_r}(\widehat{\bm{\phi}}_{\bm{x}_0}^{\bm{w}}(t))+\mathcal{L}_{\bm{h},\bm{u}}(\widehat{\bm{\phi}}_{\bm{x}_0}^{\bm{w}}(t))
\end{split}
\end{equation*}
holds for $t\in \mathbb{R}_{\geq 0}$ and $\bm{w}\in \Omega$.
Thus, we have that  for $t\in \mathbb{R}_{\geq 0}$,
\begin{equation*}
\begin{split}
\alpha \mathbb{E}[\int_{0}^t h(\widehat{\bm{\phi}}_{\bm{x}_0}^{\bm{w}}(\tau))d \tau]&\leq \alpha \mathbb{E}[\int_{0}^t 1_{\mathcal{X}_r}(\widehat{\bm{\phi}}_{\bm{x}_0}^{\bm{w}}(\tau))d\tau]\\
&+ \mathbb{E}[\int_{0}^t \mathcal{L}_{h,\bm{u}}(\widehat{\bm{\phi}}_{\bm{x}_0}^{\bm{w}}(\tau)) d\tau]
\end{split}
\end{equation*}
and thus 
\[
\begin{split}
&\alpha \int_{0}^t \mathbb{E}[h(\widehat{\bm{\phi}}_{\bm{x}_0}^{\bm{w}}(\tau))]d \tau
\\
&\leq \alpha\mathbb{E}[\int_{0}^t 1_{\mathcal{X}_r}(\widehat{\bm{\phi}}_{\bm{x}_0}^{\bm{w}}(\tau))d\tau]+ \mathbb{E}[h(\widehat{\bm{\phi}}_{\bm{x}_0}^{\bm{w}}(t))]-h(\bm{x}_0).
\end{split}
\]

Combining with \eqref{leq} we further have that 
\[
\begin{split}
\alpha h(\bm{x}_0)&\leq \alpha\frac{\mathbb{E}[\int_{0}^t 1_{\mathcal{X}_r}(\widehat{\bm{\phi}}_{\bm{x}_0}^{\bm{w}}(\tau))d\tau]}{t}\\
&+ \frac{\mathbb{E}[h(\widehat{\bm{\phi}}_{\bm{x}_0}^{\bm{w}}(t))]-h(\bm{x}_0)}{t} , \forall t\in \mathbb{R}_{\geq 0}
\end{split}
\]
and thus 
\begin{equation}
\label{ineqa11}
\begin{split}
\alpha h(\bm{x}_0)&\leq \alpha\lim_{t\rightarrow \infty}\frac{\mathbb{E}[\int_{0}^t 1_{\mathcal{X}_r}(\widehat{\bm{\phi}}_{\bm{x}_0}^{\bm{w}}(\tau))d\tau]}{t}\\
&+ \lim_{t\rightarrow \infty}\frac{\mathbb{E}[h(\widehat{\bm{\phi}}_{\bm{x}_0}^{\bm{w}}(t))]-h(\bm{x}_0)}{t}.
\end{split}
\end{equation}

Since $ \lim_{t\rightarrow \infty}\frac{\mathbb{E}[h(\widehat{\bm{\phi}}_{\bm{x}_0}^{\bm{w}}(t))]-h(\bm{x}_0)}{t}=0$ (from the compactness of the set $\overline{\mathcal{C}}$),  we have 
\[
\begin{split}
&\mathbb{P}(\{\bm{w}\in \Omega\mid \exists t\geq 0. \widehat{\bm{\phi}}_{\bm{x}_0}^{\bm{w}}(t)\in \mathcal{X}_r\})\\
&=\lim_{t\rightarrow \infty}\frac{\mathbb{E}[\int_{0}^t 1_{\mathcal{X}_r}(\widehat{\bm{\phi}}_{\bm{x}_0}^{\bm{w}}(\tau))d\tau]}{t}\geq h(\bm{x}_0).
\end{split}
\] Consequently, we have the conclusion.
\end{proof}

\textbf{The proof of Proposition \ref{s_e}:}

\begin{proof}
The conclusion can be justified by following the proof of Theorem  \ref{exponential_s} with $h(\bm{x})=h'(\bm{x})$.


\end{proof}

\textbf{The proof of Theorem \ref{asymp_s}:}

\begin{proof}
The conclusion can be assured by following arguments in Theorem \ref{exponential_s} with small modifications. 

Assume that $\bm{x}_0\in \mathcal{C}$.  According to Lemma 1 in \cite{xue2022_Stochastic}, 
\begin{equation*}
\begin{split}
    &\mathbb{P}(\{\bm{w}\in \Omega \mid \exists t\geq 0. \bm{\phi}_{\bm{x}_0}^{\bm{w}}(t) \in \mathcal{X}_r\bigwedge \forall \tau\in [0,t]. \bm{\phi}_{\bm{x}_0}^{\bm{w}}(\tau) \in \mathcal{C}\})\\
    &=\mathbb{P}(\{\bm{w}\in \Omega\mid \exists t\geq 0. \widehat{\bm{\phi}}_{\bm{x}_0}^{\bm{w}}(t) \in \mathcal{X}_r\}).
    \end{split}
\end{equation*}
Consequently, we just need to show that 
\[\mathbb{P}(\{\bm{w}\in \Omega\mid \exists t\geq 0. \widehat{\bm{\phi}}_{\bm{x}_0}^{\bm{w}}(t) \in \mathcal{X}_r\})\geq h(\bm{x}_0).\]

From constraint \eqref{stochastic_c_e_a} and Lemma 3 in \cite{xue2022_Stochastic}, we have that 
\begin{equation}
\label{leq_a}
h(\bm{x}_0)\leq \mathbb{E}[h(\widehat{\bm{\phi}}_{\bm{x}_0}^{\bm{w}}(t))], \forall t\in \mathbb{R}_{\geq 0}.
\end{equation}
Also, constraint \eqref{stochastic_c_e_a} indicates that 
\begin{equation*}
\begin{split}
 h(\widehat{\bm{\phi}}_{\bm{x}_0}^{\bm{w}}(t))\leq 1_{\mathcal{X}_r}(\widehat{\bm{\phi}}_{\bm{x}_0}^{\bm{w}}(t))+\mathcal{L}_{\bm{w},\bm{u}}(\widehat{\bm{\phi}}_{\bm{x}_0}^{\bm{w}}(t))
\end{split}
\end{equation*}
holds for $t\in \mathbb{R}_{\geq 0}$ and $\bm{w}\in \Omega$.
Thus, we have that  for $t\in \mathbb{R}_{\geq 0}$,
\begin{equation*}
\begin{split}
\mathbb{E}[\int_{0}^t h(\widehat{\bm{\phi}}_{\bm{x}_0}^{\bm{w}}(\tau))d \tau]&\leq \mathbb{E}[\int_{0}^t 1_{\mathcal{X}_r}(\widehat{\bm{\phi}}_{\bm{x}_0}^{\bm{w}}(\tau))d\tau]\\
&+ \mathbb{E}[\int_{0}^t \mathcal{L}_{w,\bm{u}}(\widehat{\bm{\phi}}_{\bm{x}_0}^{\bm{w}}(\tau)) d\tau]
\end{split}
\end{equation*}
and thus 
\[
\begin{split}
&\int_{0}^t \mathbb{E}[h(\widehat{\bm{\phi}}_{\bm{x}_0}^{\bm{w}}(\tau))]d \tau
\\
&\leq \mathbb{E}[\int_{0}^t 1_{\mathcal{X}_r}(\widehat{\bm{\phi}}_{\bm{x}_0}^{\bm{w}}(\tau))d\tau]+ \mathbb{E}[w(\widehat{\bm{\phi}}_{\bm{x}_0}^{\bm{w}}(t))]-w(\bm{x}_0).
\end{split}
\]

Combining with \eqref{leq_a} we further have that 
\[
\begin{split}
h(\bm{x}_0)&\leq \frac{\mathbb{E}[\int_{0}^t 1_{\mathcal{X}_r}(\widehat{\bm{\phi}}_{\bm{x}_0}^{\bm{w}}(\tau))d\tau]}{t}\\
&+ \frac{\mathbb{E}[w(\widehat{\bm{\phi}}_{\bm{x}_0}^{\bm{w}}(t))]-w(\bm{x}_0)}{t} , \forall t\in \mathbb{R}_{\geq 0}
\end{split}
\]
and thus 
\begin{equation}
\label{ineqa11_a}
\begin{split}
h(\bm{x}_0)&\leq \lim_{t\rightarrow \infty}\frac{\mathbb{E}[\int_{0}^t 1_{\mathcal{X}_r}(\widehat{\bm{\phi}}_{\bm{x}_0}^{\bm{w}}(\tau))d\tau]}{t}\\
&+ \lim_{t\rightarrow \infty}\frac{\mathbb{E}[w(\widehat{\bm{\phi}}_{\bm{x}_0}^{\bm{w}}(t))]-w(\bm{x}_0)}{t}, \forall t\in \mathbb{R}_{\geq 0}.
\end{split}
\end{equation}

Since $ \lim_{t\rightarrow \infty}\frac{\mathbb{E}[w(\widehat{\bm{\phi}}_{\bm{x}_0}^{\bm{w}}(t))]-w(\bm{x}_0)}{t}=0$,  we have 
\[
\begin{split}
&\mathbb{P}(\{\bm{w}\in \Omega\mid \exists t\geq 0. \widehat{\bm{\phi}}_{\bm{x}_0}^{\bm{w}}(t)\in \mathcal{X}_r\})\\
&=\lim_{t\rightarrow \infty}\frac{\mathbb{E}[\int_{0}^t 1_{\mathcal{X}_r}(\widehat{\bm{\phi}}_{\bm{x}_0}^{\bm{w}}(\tau))d\tau]}{t}\geq h(\bm{x}_0).
\end{split}\] Consequently, we have the conclusion.
\end{proof}